\documentclass[letterpaper,twocolumn,10pt]{article}
\usepackage{zhanggroup}

\usepackage{booktabs}
\usepackage{multirow}
\usepackage{amsthm}
\usepackage{amsmath}
\usepackage{amssymb}
\usepackage{xspace}
\usepackage{xcolor}
\usepackage{colortbl}
\usepackage{graphicx}
\usepackage{caption,subcaption}
\usepackage{picture,graphics}
\usepackage[absolute]{textpos}
\DeclareMathOperator*{\argmax}{arg\,max}

\newcommand{\lm}{\ensuremath{\mathcal{M}\xspace}}
\newcommand{\prompt}{\ensuremath{\texttt{prompt}\xspace}}
\newcommand{\instruct}{\ensuremath{\textit{I}\xspace}}
\newcommand{\refappendix}[1]{\hyperref[#1]{Appendix~\ref*{#1}}}

\theoremstyle{definition}

\theoremstyle{remark}

\definecolor{mygray}{gray}{.9}
\newcommand{\mypara}[1]{\noindent\textbf{#1:}}

\begin{document}

\begin{textblock}{15}(1.9,1)
To Appear in 2024 ACM SIGSAC Conference on Computer and Communications Security, October 14-18, 2024
\end{textblock}

\date{}

\title{\Large \bf Membership Inference Attacks Against In-Context Learning} 

\author{
{\rm Rui Wen}\ \ \
{\rm Zheng Li\thanks{Corresponding authors.}}\ \ \
{\rm Michael Backes}\ \ \
{\rm Yang Zhang\thanksmark{1}}\ \ \
\\
\\
\textit{CISPA Helmholtz Center for Information Security}\\
}

\maketitle

\begin{abstract}
Adapting Large Language Models (LLMs) to specific tasks introduces concerns about computational efficiency, prompting an exploration of efficient methods such as In-Context Learning (ICL). 
However, the vulnerability of ICL to privacy attacks under realistic assumptions remains largely unexplored.
In this work, we present the first membership inference attack tailored for ICL, relying solely on generated texts without their associated probabilities. 
We propose four attack strategies tailored to various constrained scenarios and conduct extensive experiments on four popular large language models.
Empirical results show that our attacks can accurately determine membership status in most cases, e.g., 95\% accuracy advantage against LLaMA, indicating that the associated risks are much higher than those shown by existing probability-based attacks.
Additionally, we propose a hybrid attack that synthesizes the strengths of the aforementioned strategies, achieving an accuracy advantage of over 95\% in most cases.
Furthermore, we investigate three potential defenses targeting data, instruction, and output. 
Results demonstrate combining defenses from orthogonal dimensions significantly reduces privacy leakage and offers enhanced privacy assurances.
\end{abstract}

%-------------------------------------------------------------------------------
\section{Introduction}
\label{sec:intro}
%-------------------------------------------------------------------------------

The rapid evolution of Large Language Models (LLMs) has garnered widespread attention, reshaping various facets of contemporary society. 
These models, distinguished by their remarkable capabilities, have been instrumental in augmenting diverse aspects of human life. 
However, customizing LLMs for specific domains often involves computationally inefficient adjustments.

To overcome these challenges, In-Context Learning (ICL)~\cite{BMRSKDNSSAAHKHCRZWWHCSLGCCBMRSA20} emerges as a novel and efficient approach to task-specific adaptation within the paradigm of LLMs. 
Unlike conventional fine-tuning, ICL does not necessitate extensive updates to model parameters. 
Instead, it harnesses the power of prompts—additional contextual content—to guide the model's learning through analogy~\cite{DLDZWCSXLS23}.

While ICL offers substantial advantages, its integration into language models raises a critical issue:  vulnerability to privacy breaches. 
This concern is particularly pronounced for language models designed for personalization and adaptation to user-specific inputs, where the prompt becomes a repository of sensitive information. 
For instance, LLMs integrated with ICL are increasingly deployed in healthcare analytics~\cite{MRKPBG23,SGBCGSP23,MHWYZDRRL23}.
Knowing that the victim's data belongs to the model's training data, i.e., the prompt of the ICL, the adversary can immediately learn about the victim's health status.
This form of sensitive information leakage stems from one of the most serious privacy threats in machine learning, namely Membership Inference Attacks (MIAs)~\cite{NSH19,LF20,SSSS17,SZHBFB19,LLR21,SM21,LZ21,SCEKPSTB23}.

In this case, MIAs aim to determine whether a data sample has been used for in-context learning, and their success holds two significant implications. 
Firstly, MIAs represent a fundamental form of a privacy attack, offering insights into more sophisticated attacks and implying diverse privacy vulnerabilities~\cite{SCEKPSTB23}.
Secondly, MIAs can serve as a valuable tool for auditing data provenance. 
Existing MIAs targeting language models rely on the probability associated with the texts returned by LLMs~\cite{DDYPB23,FWGLLJ23,SAXHLBCZ23,MMJSSB23}.
However, a significant drawback of these probability-based attacks is that they can be easily mitigated if LLMs only return generated texts (which is actually the current realistic scenario).
The fact that probability-based attacks can be easily averted makes it more difficult to evaluate whether a model is truly vulnerable to membership inference or not, which may lead to premature claims of privacy for LLMs.

In this work, we concentrate on the membership leakage of in-context learning, and present the first text-only membership inference attack that relies only on the final text generated by the language model.
Specifically, we propose four attack methods: GAP, Inquiry, Repeat, and Brainwash. 
The GAP attack serves as a baseline, considering samples as members if correctly classified and as non-members if not. 
Among the three advanced attacks, the Inquiry attack directly asks the language model whether it has encountered specific samples. 
The Repeat attack identifies samples as members if the language model can generate text that closely matches the original input. 
Finally, in more challenging scenarios where the model produces fixed responses like ``positive'' or ``negative,'' we introduce the Brainwash attack. 
This novel method consistently influences the model to provide specific incorrect answers, and membership is inferred based on the sample's ability to conform to this brainwashing process.

We conduct extensive experiments on four popular large language models and three benchmark datasets.
Empirical results show that our attacks achieve significant performance across various scenarios.
For example, the Brainwash attack achieves over 95\% accuracy advantage in inferring membership status against LLaMA on the DBPedia and AGNews datasets.
Even when applied to online commercial models such as GPT-3.5, our attack maintains an advantage of over 60\% on the TREC dataset.

We further comprehensively investigate factors influencing the attack, including the number of demonstrations and their position in the prompt. 
Results indicate that the vulnerability of demonstrations emerges from a synergistic interplay between prompt size and the demonstration position. 
These findings offer insights for designing prompts that are more resilient against privacy attacks.
Moreover, we undertake a comprehensive case study investigating the evolving behavior of updated language models over time. 
Despite ongoing efforts to enhance model safety, our results reveal persistent vulnerabilities in the prompt, even with updated versions.

Recognizing the applicability of our attacks across diverse scenarios, we further design a hybrid attack that combines the strengths of the Brainwash and Repeat attacks. 
Empirical evidence shows that the hybrid attack outperforms both individual methods in most cases, e.g., 81.2\% accuracy advantage compared to 67.8\% and 73.0\%, respectively.
Lastly, we explore three potential defenses at the data, instruction, and output levels. 
Our results demonstrate that these defenses are effective for specific attacks and datasets. 
Additionally, we find that combining defenses from all these orthogonal dimensions significantly mitigates privacy leakage and provides stronger privacy guarantees.

We summarize our contribution as follows:
\begin{itemize}
    \item We present the first text-only membership inference attack against ICL. 
    We design four text-only attacks and empirically demonstrate their effectiveness across four popular language models and three diverse datasets.
    \item We conduct extensive studies of factors influencing attack performance and reveal that the vulnerability of demonstrations emerges as a synergistic interplay between prompt size and the demonstration position.
    \item We integrate two powerful attacks to construct a hybrid attack, which significantly enhances our attack's performance and generalizability.
    \item We explore three potential defenses for ICL against text-only attacks and empirically show their effectiveness in mitigating privacy leakage.
\end{itemize}

%-------------------------------------------------------------------------------
\section{Preliminaries}
\label{sec:pre}
%-------------------------------------------------------------------------------

%-------------------------------------------------------------------------------
\subsection{In-Context Learning}
\label{sec:pre_icl}
%-------------------------------------------------------------------------------

\begin{figure}[!t]
\centering
\includegraphics[width=0.88\columnwidth]{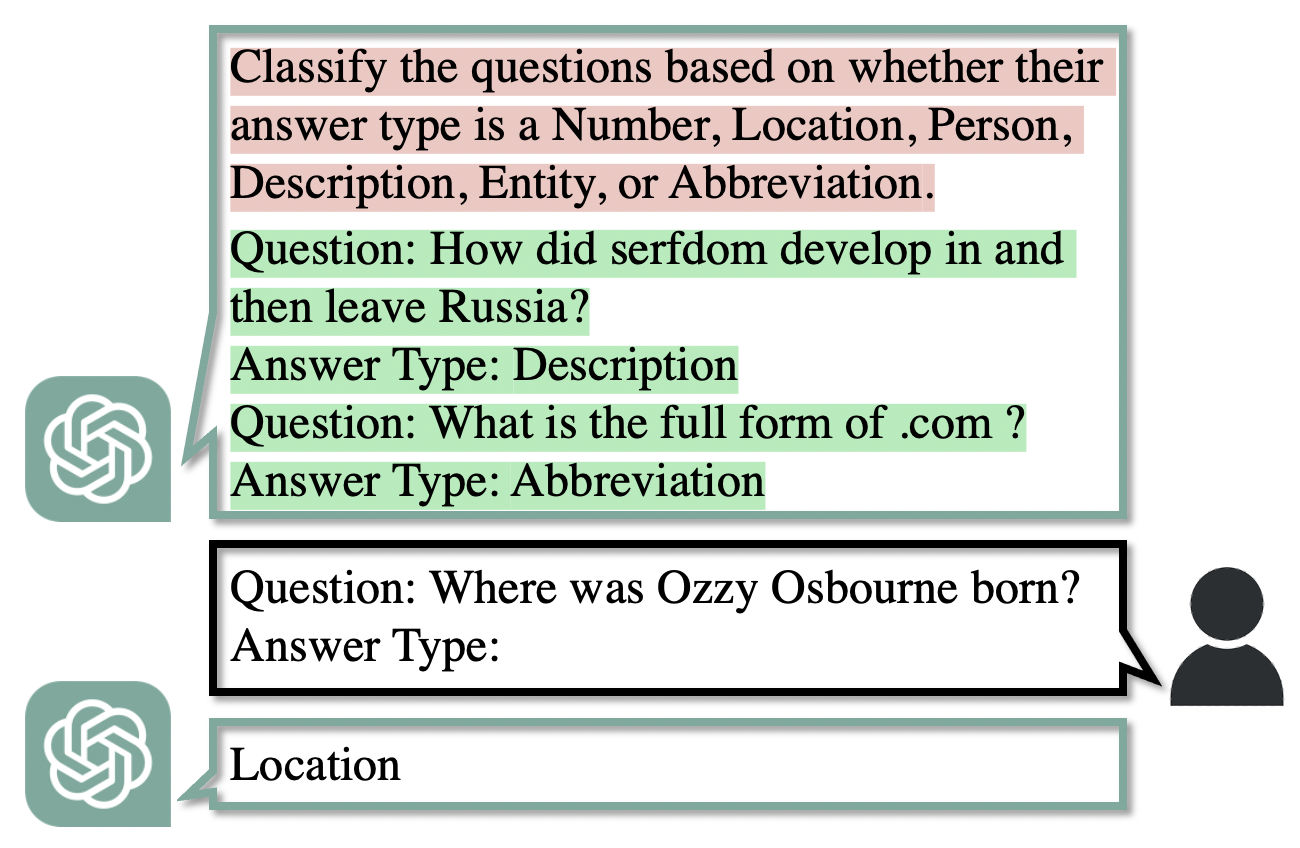}
\caption{An illustrative example of In-Context Learning.
The language model is initialized by a prompt combined with instruction (pink) and demonstrations (green).}
\label{figure:icl_demo}
\end{figure}

In-Context Learning (ICL) emerges as a distinctive feature of large language models (LLMs)~\cite{BMRSKDNSSAAHKHCRZWWHCSLGCCBMRSA20}, affording LLMs the capability to acquire proficiency in specific tasks through exposure to limited demonstration examples. 
In contrast to the conventional notion of ``learning,'' In-Context Learning does not require updating model parameters. 
Instead, it augments the input with additional content, referred to as a prompt, to facilitate learning through analogy~\cite{DLDZWCSXLS23}.
Specifically, within the prompt, the model is provided with several input-output pairs as examples, instructing the model to respond in a similar format.

To integrate ICL into LLMs, the model undergoes an initialization process. 
This process involves carefully constructing a task-specific prompt, encompassing an optional task instruction (\instruct) and $k$ demonstration examples ($\{(x_i,y_i)|i\leq k, i\in\mathbb{N}^+\}$).
The server concatenates these components to form a complete prompt, denoted as $\prompt = \{\instruct, s(x_1, y_1),\dots, s(x_k, y_k)\}$. 
Here, the function $s(\cdot,\cdot)$ denotes the transformation of demonstration pairs into natural language following a predefined template.
In addition, for a certain task in ICL, the number of demonstrations is not large, typically no more than 8, i.e., $k\leq8$. 
This is due to a trade-off between input size and performance. 
Increasing the number of demonstrations beyond eight results in only marginal performance improvements~\cite{ZWFKS21}.
We provide an illustrative case in ~\autoref{figure:icl_demo}.
Highlighted in pink are the task instructions, which instruct the language model to classify the questions into different categories, and in green are the two demonstrations.

During testing, the language model accepts input samples $x$ in the same format as the $\prompt$ demonstration, i.e., \textit{``Question: x; Answer Type:\{ \}''}.
Subsequently, the model assigns probabilities $P(y_i|x,\prompt)$ to all potential answers $y_i\in \mathcal{Y}$, and selects the output token based on sampling strategies, such as greedy decoding, which selects the token with the highest probability. 
Mathematically, this process is represented as:
\[
\argmax_{y_i\in \mathcal{Y}}P(y_i|x,\prompt).
\]

It is important to note that in ICL, the term ``model's training data'' might be misleading. 
While the prompt contains demonstration data used to guide the model's responses, there is no actual retraining of the model's weights involved. 
Instead, the model uses the provided examples to draw analogies and make predictions, simulating a form of learning without altering its underlying parameters. 
This distinction is crucial for understanding the model's behavior and the potential vulnerabilities associated with ICL.

%-------------------------------------------------------------------------------
\subsection{Membership Inference Attack}
\label{sec:pre_mia}
%-------------------------------------------------------------------------------

Membership inference attack (MIA) represents one of the most basic forms of privacy attack~\cite{SCEKPSTB23}, where the attacker aims to determine whether a given sample belongs to the training dataset.
Such an attack has been widely studied in the traditional ML domain, as leaking the membership information could cause several implications.
This concern persists in the realm of Large Language Models (LLMs), where the revelation of the $\prompt$ containing specific data translates to a breach of private information. 
This becomes particularly consequential in sensitive tasks. 
Furthermore, the significance of MIA actually extends beyond its primary privacy implications.
Specifically, knowing the $\prompt$ used by an LLM allows an adversary to obtain extra information about the LLM's ICL prompts, which infringes on the LLM's intellectual property rights.
From another perspective, MIA can be used as an auditing tool by users to find out whether their data is being collected to craft the LLM's ICL prompts.

The theoretical rationale behind existing membership inference attacks is mainly based on the observation that models exhibit varying degrees of confidence in their responses, particularly favoring samples encountered during training. 
Membership inference attacks employ a variety of approaches, one prominent and straightforward method is to use the posteriors to train an attack model. 
In this approach, samples with more ``member-like'' posteriors are classified as members. 
Additionally, efforts to enhance MIA performance have been explored by incorporating additional information such as model intermediate representation~\cite{NSH19} or loss trajectory~\cite{LZBZ22}, and by actively training shadow models with carefully crafted datasets~\cite{CCNSTT22}. 
In all of these instances, it seems that having access to the model posterior is a necessary requirement for launching the attack.

Recent research~\cite{CTCP21,LZ21} endeavors have delved into the prospect of attacking models without direct access to the posterior. 
These efforts leverage the distance between the target sample and the decision boundary to predict membership status. 
However, the opacity of model architecture/parameters and the discrete input space pose challenges for extending this approach to large language models.

To the best of our knowledge, all existing membership inference attacks against LLMs necessitate, at a minimum, access to the probability associated with predictions. 
This requirement is crucial for calculating corresponding loss~\cite{WWBZS23,DDYPB23} or perplexity~\cite{CLEKS19,CTWJHLRBSEOR21}, which can then be used to extract membership signals. 
In this work, we explore the most stringent scenarios where only the generated content/text is available to the adversary.
We refer to such kind of attack as \textit{text-only} membership inference attack.
Besides, we emphasize here that compared to probability-based attacks, text-only attacks are much more realistic in real-world applications where probabilities are rarely accessible.

%-------------------------------------------------------------------------------
\section{Problem Statement}
\label{sec:problem}
%-------------------------------------------------------------------------------

\mypara{Adversary’s Objective}
The primary objective of the adversary is to determine whether a specific target sample $x$ was included in the construction of a prompt used to customize a language model $\lm$. 
The prompt, denoted as $\prompt$, comprises a set of $k$ demonstrations, formatted as $\prompt = \{\instruct, s(x_1, y_1),\dots, s(x_k, y_k)\}$. 
The adversary's goal is to determine whether the target sample $x$ has been utilized in crafting the $\prompt$.
That is, the goal is to determine whether $x$ is in the set $\{x_1,\dots,x_k\}$.

\mypara{Adversary’s Capabilities}
In this work, the adversary can access tailored language models that are customized via prompts with \textit{fixed} demonstrations,  as indicated in previous research~\cite{ZLWJZBSZ24,TSIMMLGKS23}. 
For example, Copy.ai~\cite{CopyAI} suggests employing well-crafted prompts that include examples and stylistic instructions to generate high-quality marketing copy, and these prompts do not change between requests. 
Furthermore, as GPTs~\cite{GPTs} (powered by GPT-3.5/4) and GLMs~\cite{GLMs} (powered by ChatGLM4) gain increasing interest, we anticipate more use cases of customized LLMs with fixed demonstrations to perform user-determined tasks, such as sentiment analysis~\cite{WXDFX23} and text summarization~\cite{JKSFLNZ23}.

More concretely, we consider the most strict and realistic scenario where the adversary has only black-box access to the target language model $\lm$, meaning they can see the text generated but \textit{not} the tokenizer or associated probabilities. 
Additionally, the adversary has the ground truth answer $y$ for the target sample $x$. 
This assumption aligns with most existing membership inference works in computer vision~\cite{YMMBS22, HSSDYZ21,LLHYBZ22} and natural language processing domains~\cite{DDYPB23,WWBZS23,MMJSSB23}.

\begin{figure}[t]
\centering
\includegraphics[width=0.86\columnwidth]{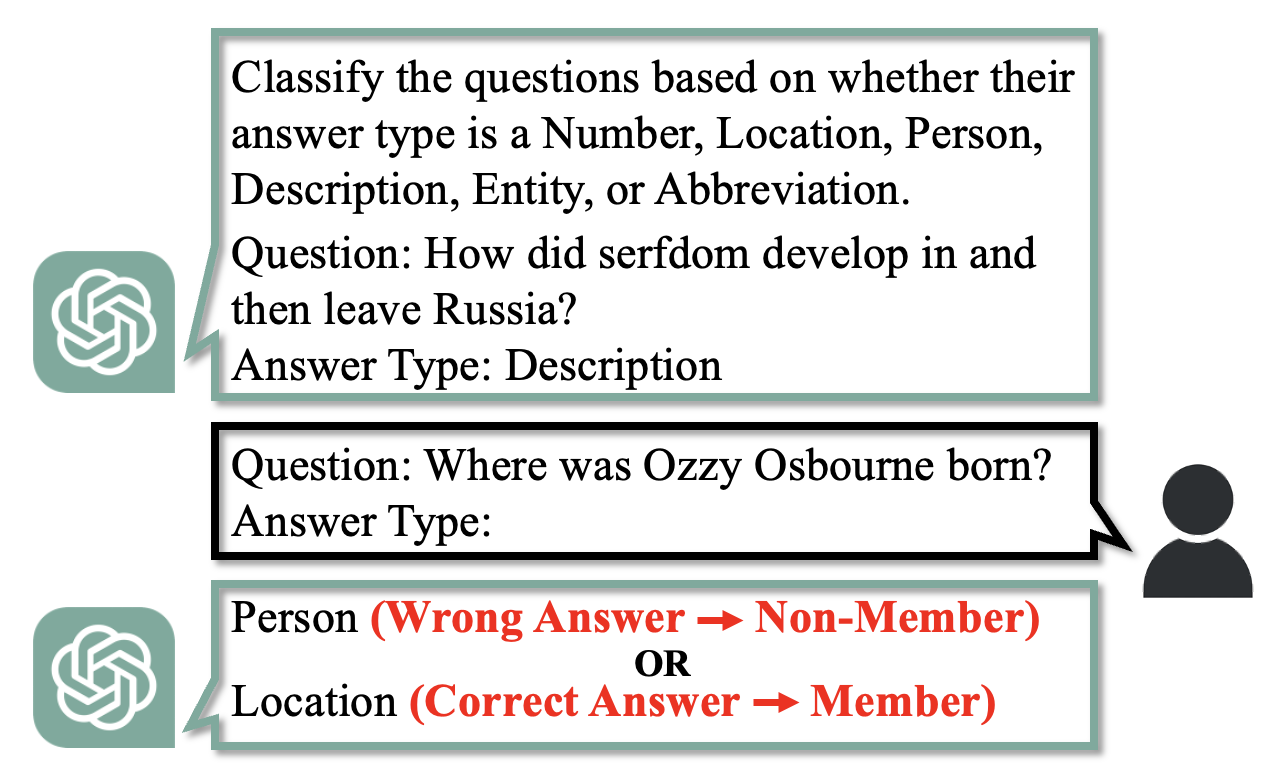}
\caption{The GAP attack involves querying the model with a target sample. 
The adversary determines the membership status by evaluating the accuracy of the model's prediction: if the prediction is correct, the target sample is classified as a member; otherwise, it is classified as a non-member.}
\label{figure:gap_illus_workflow}
\end{figure}

%-------------------------------------------------------------------------------
\section{Attack Methodology}
%-------------------------------------------------------------------------------

%-------------------------------------------------------------------------------
\subsection{A Baseline Attack: GAP Attack}
\label{sec:vanilla}
%-------------------------------------------------------------------------------

We start with a baseline attack that extends from the existing attack in the vision domain.
Under the assumption that the adversary only has access to the generated texts without additional details, a straightforward approach for membership inference involves exploiting the well-known overfitting phenomenon, where models tend to memorize samples from the training dataset, thereby exhibiting higher accuracy on these than on the testing dataset.

Prior research~\cite{RWAACBS19} indicates language models exhibit minimal overfitting due to their massive training dataset and fewer training epochs on individual data points~\cite{RSRLNMZLL20}. 
However, In-Context Learning introduces a potential vulnerability, as it allows language models to recall recently encountered demonstrations, thus behaving as if they have ``memorized'' them.

Building on this, we categorize samples that are correctly identified as ``members'' and the rest as ``non-members.'' 
This approach is an uncomplicated extension of existing work in the vision domain~\cite{YGFJ18} to language model settings. 
We refer to this basic attack methodology as the GAP attack, which serves as our starting point and baseline for comparison.

\mypara{Methodology} 
The attack methodology is structured as follows (see~\autoref{figure:gap_illus_workflow} for an illustration):
\begin{itemize}
    \item The adversary selects a target sample $x$, which is a sentence whose membership status they aim to determine.
    \item The adversary sends the target sample $x$ to the model and observes the model’s response. 
    If the model returns the correct answer, the sentence is classified as a member of the dataset; if not, it is deemed a non-member.
\end{itemize}

\begin{figure}[!t]
\centering
\includegraphics[width=0.66\columnwidth]{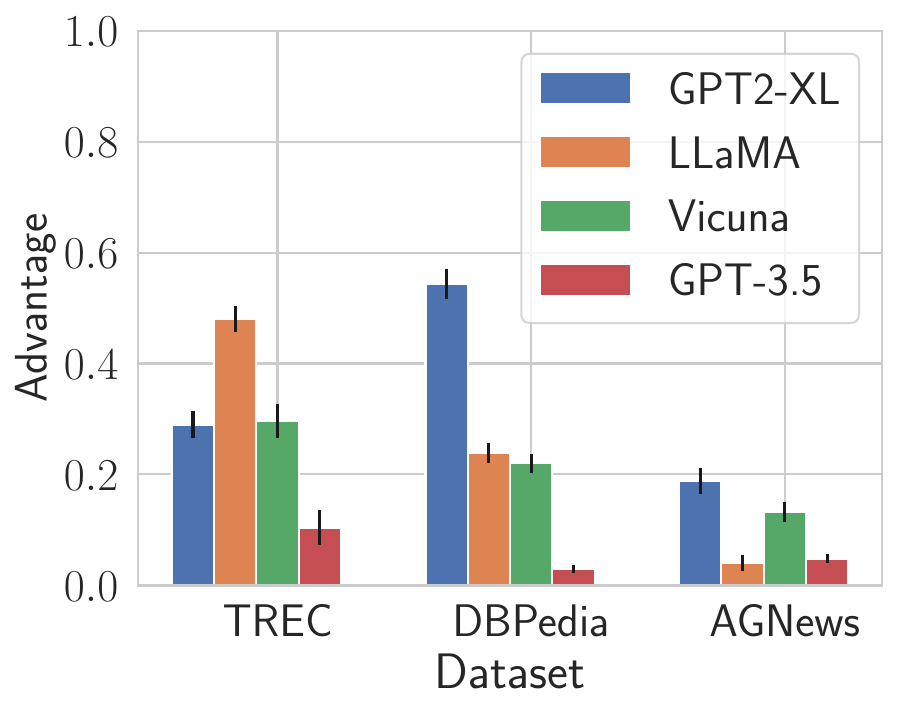}
\caption{Performance of the baseline membership inference attack (GAP), revealing challenges and suboptimal results, particularly evident in larger language models such as GPT-3.5. 
In this figure, language models are prompted with one example with the instruction presented in~\autoref{figure:gap_illus_workflow}. 
The performance metric, which indicates the advantage over random guessing, is detailed in~\autoref{sec:setup}.}
\label{figure:gap_illus}
\end{figure}

The results, illustrated in ~\autoref{figure:gap_illus}, reveal unsatisfactory performance, particularly for LLMs like GPT-3.5 (0 means random guess), as these models perform very well even when the test samples are not seen in the prompt. 
This suboptimal performance motivates us to develop more effective attacks tailored specifically for LLMs.

\begin{figure}[!t]
\centering
\includegraphics[width=0.86\columnwidth]{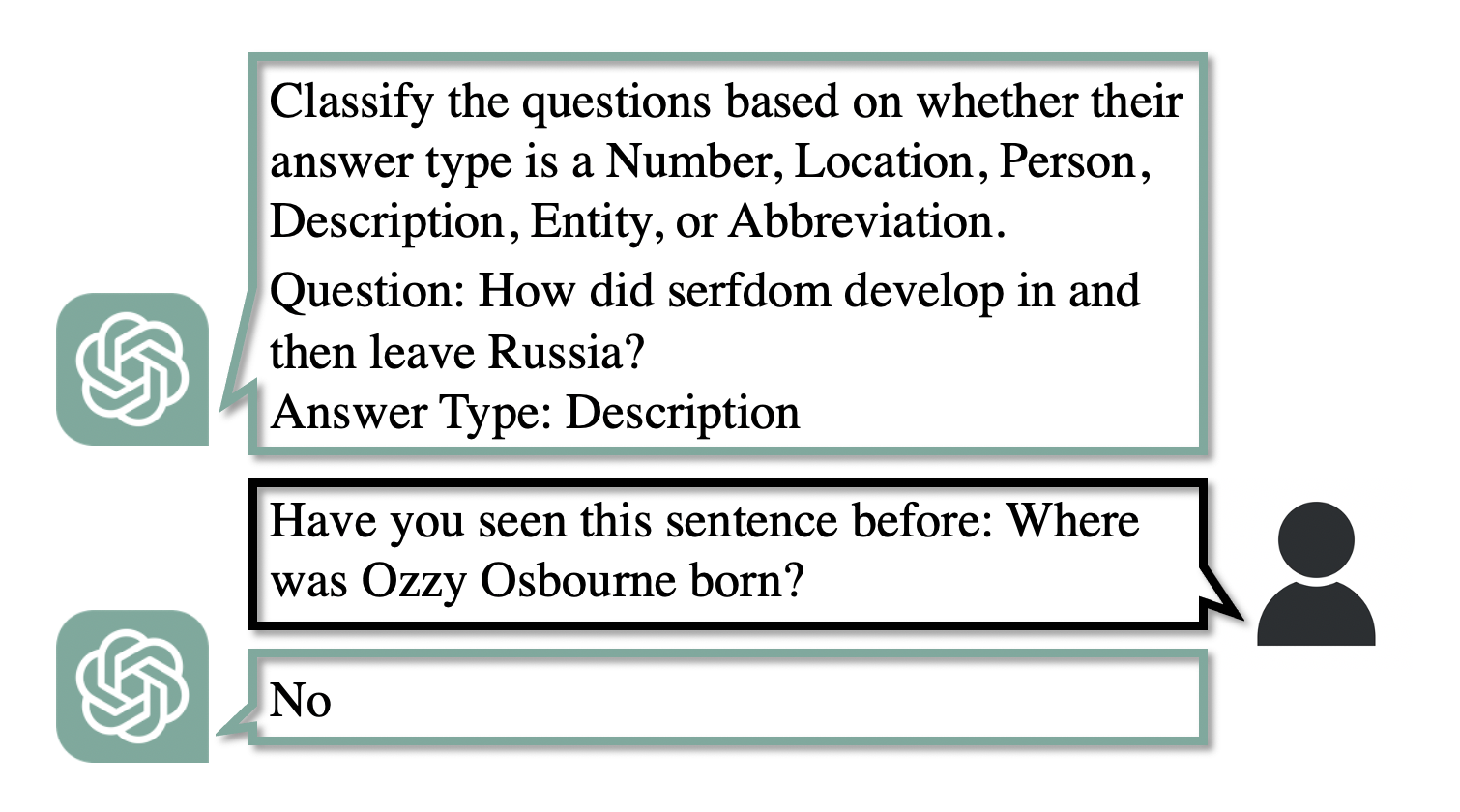}
\caption{The Inquiry attack determines membership status by directly querying the model. 
In our work, we use the prompt ``Have you seen this sentence before.''}
\label{fig:attack_inquiry}
\end{figure}

%-------------------------------------------------------------------------------
\subsection{Inquiry Attack}
%-------------------------------------------------------------------------------

\mypara{Intuition} 
The core concept of this attack method hinges on the language model's ability to remember information from past conversations and deliver context-based responses. 
When we interact with a language model, it processes the context and produces a response informed by the knowledge it has acquired from previous inputs by the user, particularly from the provided prompt ($\prompt$) and its included demonstrations. 
Consequently, a direct and intuitive approach is to directly question the language model about its previous encounters with specific samples.

\mypara{Methodology} 
The attack methodology is structured as follows (refer to~\autoref{fig:attack_inquiry} for an illustration):
\begin{itemize}
    \item The adversary selects a target sample $x$, which is a sentence that they aim to determine its membership status.
    \item The adversary crafts a query to the model with the prompt: \textit{``Have you seen this sentence before: \{$x$\}?''} 
    \item The adversary sends the query to the model and observes the model’s response. 
    If the model confirms with a ``yes'', the sentence is classified as a member of the dataset; if not, it is deemed a non-member.
\end{itemize}

%-------------------------------------------------------------------------------
\subsection{Repeat Attack}
%-------------------------------------------------------------------------------

The Inquiry attack, while direct and straightforward, may trigger alerts and consequently be denied a response as it overtly queries the language model's prompt.
To mitigate this risk, we introduce the Repeat attack, which employs a more subtle approach.

\mypara{Intuition} 
This attack leverages the strong memorization capability of language models to generate context-aware responses. 
Unlike the Inquiry attack, which uses queries with clear intentions (such as \textit{"Have you seen this sentence before?"}), we use the core functionality of a language model, which is to predict the next words. 
When provided with just the beginning of a target sample, such as the first three words, the model attempts to complete the sentence by adding more words. 
Our hypothesis is that the model's prior knowledge, enhanced through ICL, will encourage the language model to generate text that mirrors previously encountered content.

\begin{figure}[!t]
\centering
\includegraphics[width=0.86\columnwidth]{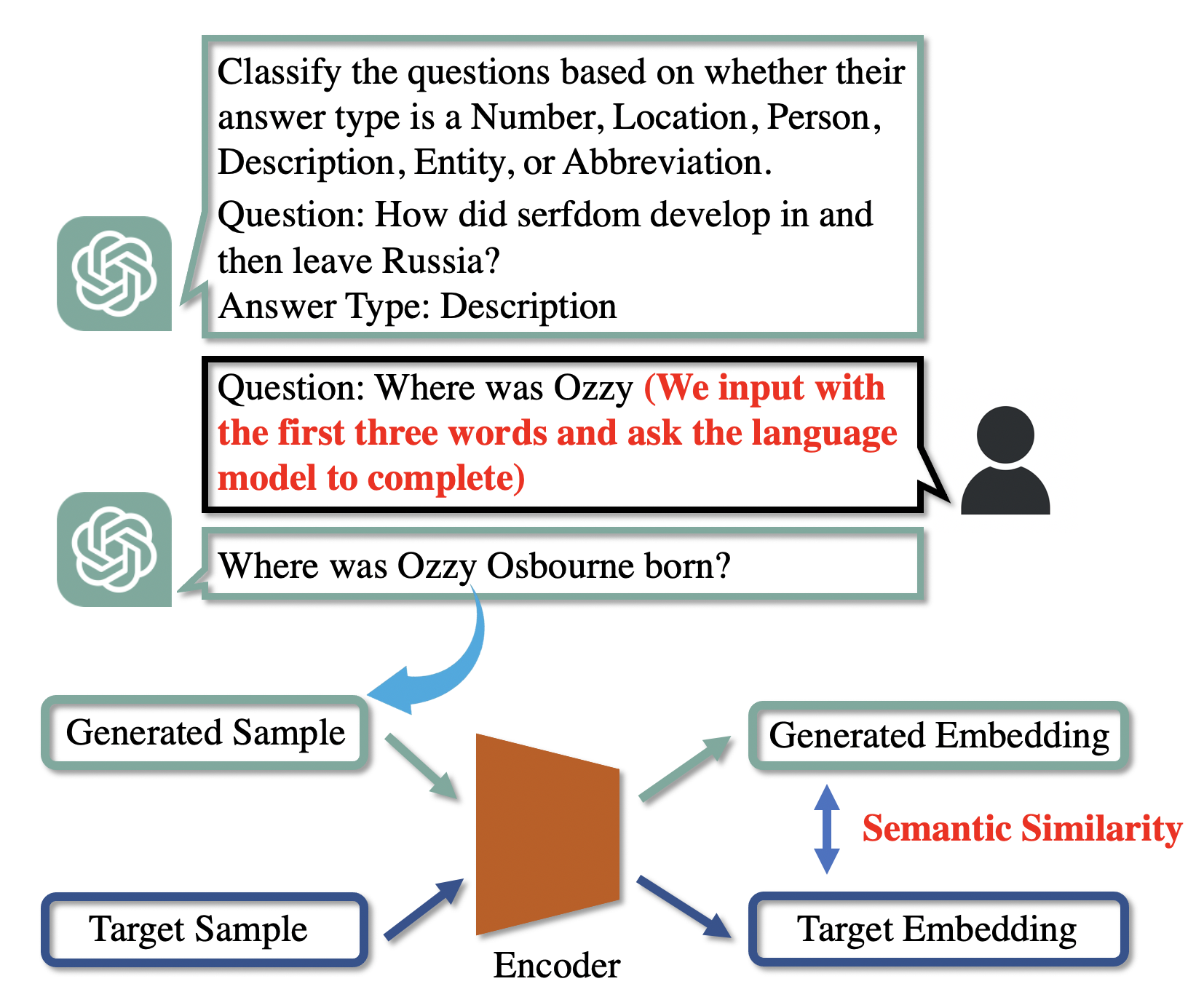}
\caption{The Repeat attack initiates a conversation with a few words and asks the model to complete the sentence. 
The adversary predicts membership status by assessing the semantic similarity between the generated sample and the target sample.}
\label{fig:attack_repeat}
\end{figure}

\mypara{Methodology} The attack method consists of the following steps (see \autoref{fig:attack_repeat} for an illustration):
\begin{itemize}
    \item The adversary selects a target sample $x$, which is a sentence that they aim to determine its membership status.
    \item The adversary truncates the target sample, retaining only its first few words, which are then inputted to the language model. 
    The generated response $x^{'}$ from the model is obtained. 
    \item The adversary then feeds $x$ and $x^{'}$ to a text encoder \textit{E} to extract their embeddings and measure the semantic similarity between them by a function $\Phi$. 
\begin{eqnarray}
\textit{Similarity} & = & \Phi (\textit{E}(x), \textit{E}(x^{'}))
\end{eqnarray}
    If the similarity score exceeds a predetermined threshold, the sample is classified as a member of the dataset; otherwise, it is classified as a non-member.
\end{itemize}

\begin{figure}[!b]
\centering
\includegraphics[width=0.66\columnwidth]{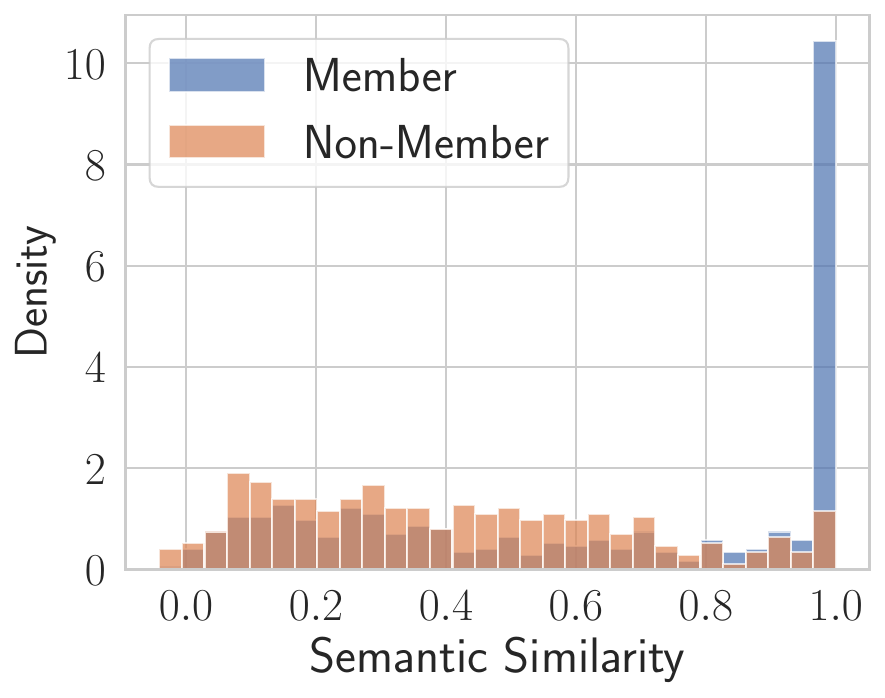}
\caption{Member samples are more likely to exhibit high similarity with the generated sample, while the similarity distribution for non-members is more flattened. 
These distributions were obtained by querying the GPT-3.5 model using the procedure detailed in~\autoref{fig:attack_repeat} with the TREC dataset, using one example as a demonstration.}
\label{figure:repeat_illus}
\end{figure}

For practical implementation, we use the first three words of the sentence to prompt the language model. 
The SentenceTransformer network~\cite{RG19}, a widely utilized text encoder, is employed to calculate semantic similarity using cosine distance.
~\autoref{figure:repeat_illus} shows that member samples generally exhibit higher similarity with the generated text, while non-member samples display a more flattened similarity distribution. 
This supports our initial hypothesis.

In practice, setting a similarity threshold between 0.8 and 0.9 tends to produce satisfactory accuracy. 
Further fine-tuning of this threshold through additional sample training can enhance the effectiveness of the attack.

%-------------------------------------------------------------------------------
\subsection{Brainwash Attack}
%-------------------------------------------------------------------------------

In this scenario, we explore a more common and strict scenario where the language model's output is confined to a predefined list of responses. 
This does \textit{not} entail modifying the model’s basic operational framework, like converting it into a classification model based on Large Language Models. 
Instead, the language model continues to generate responses in an autoregressive manner \textit{as before}. 
The key difference here is the introduction of a server-side filter, which evaluates the outputs to ensure they are permissible and non-harmful. 
This added layer heightens the complexity of launching effective attacks.

For instance, in the sentiment analysis task, the language model is limited to outputting ``positive'' or ``negative'' predictions. 
However, to the user, the interaction with the model remains unchanged.

In such a controlled output environment, launching attacks like the GAP attack remains feasible, though it represents a suboptimal method. 
More sophisticated attacks, such as those that depend on inducing the model to repeat sentences or to disclose previously seen sentences, are unlikely to succeed due to the server's filtering mechanism.

To overcome this challenge, we present the last text-only membership inference attack, namely Brainwash Attack. 
This text-only membership inference attack is designed to operate effectively under the stringent conditions imposed by output filters.

\begin{figure}[!t]
\centering
\includegraphics[width=0.86\columnwidth]{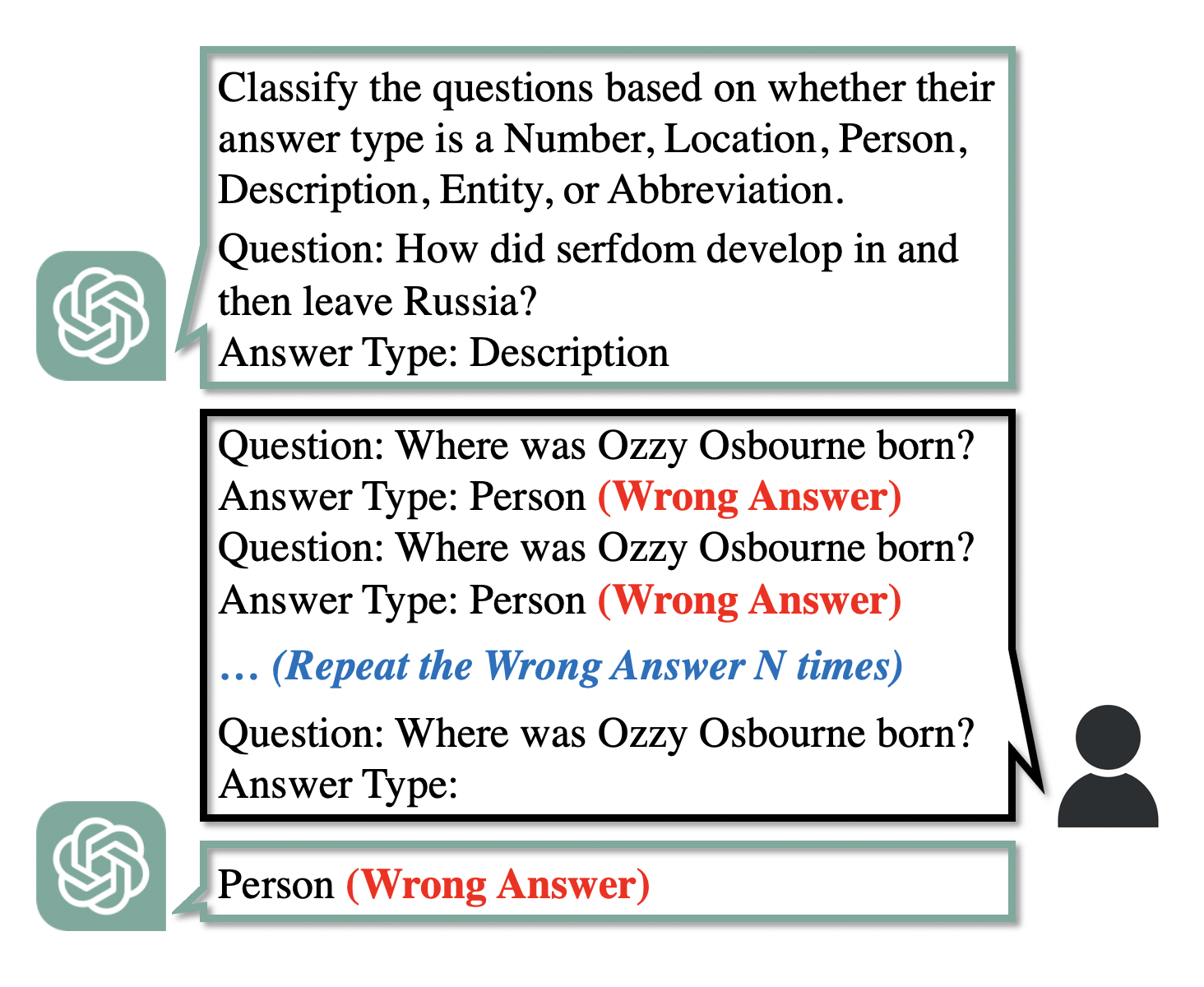}
\caption{The Brainwash attack persistently presents the target sample to the model with a consistent incorrect answer until the model responds inaccurately. 
The number of iterations required indicates the likelihood of membership.}
\label{fig:attack_brainwash}
\end{figure}

\mypara{Intuition}
Initially, let's consider a simplified case where an adversary has access to the probability or logits associated with outputs. 
Under these conditions, determining membership is straightforward: a higher probability suggests that the model is more confident in its prediction, likely because the item was seen in the prompt.
The main idea of this attack is to approximate the confidence, which is challenging given the limitations of the language model output.

To address this challenge, we approximate confidence by evaluating how firm the model is on previously encountered correct answers when it is ``brainwashed'' by unreasonable or incorrect queries. 
Specifically, if an incorrect fact is presented to the model and it hasn't encountered this information before, the model is more susceptible to being misled. 
Conversely, if the model is familiar with the correct information from its prompt, it is less likely to accept the incorrect fact.

\begin{figure}[!t]
\centering
\includegraphics[width=0.66\columnwidth]{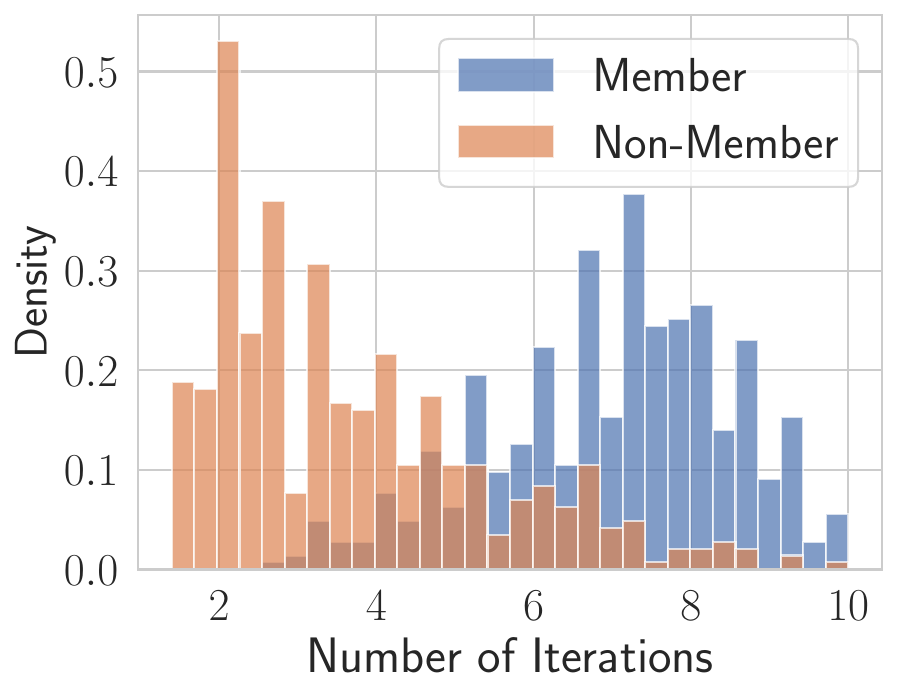}
\caption{Member samples resist incorrect labels, requiring more iterations to change the model's output, while non-members are more easily influenced.
These distributions were obtained by querying the GPT-3.5 model using the procedure detailed in~\autoref{fig:attack_brainwash} with the TREC dataset, using one example as a demonstration.}
\label{figure:brainwash_illus}
\end{figure}

\mypara{Methodology} The approach involves several key steps, as outlined below and illustrated in~\autoref{fig:attack_brainwash}:
\begin{itemize}
    \item The adversary selects a target sample $x$, which is a sentence that they aim to determine its membership status. 
    In addition, the adversary knows its correct answer $y$.
    \item The adversary crafts a query to the model with the same template, for example: \textit{``Question: x; Answer Type: $\hat{y}$.''} Here, $\hat{y}$ denotes the wrong answer compared to the correct answer $y$.
    \item The adversary repeats querying the model with the above prompt until the model responds with the incorrect answer $\hat{y}$. 
    \item The attacker then counts the number of queries needed for the model to accept the incorrect answer. 
    If this number exceeds a predefined threshold, the sample is classified as a member; otherwise, as a non-member.
\end{itemize}
In our experiments, we consider multiple choice for the incorrect answer $\hat{y}$.
The adversary counts the number of queries for each incorrect answer, and we use the average number of queries as a robust metric for evaluating confidence.
\autoref{figure:brainwash_illus} demonstrates that member samples necessitate significantly more queries to output the incorrect answer, i.e., the language model is much firmer for the correct answer it has seen before.
In contrast, non-member samples are more likely to be influenced to output incorrect answers.
This observation confirms our intuition.
We empirically determined that setting the threshold between 3 to 4 is a reasonable choice for most models and datasets, although the optimal threshold selection necessitates a few additional samples for refinement.

%-------------------------------------------------------------------------------
\section{Experiments}
%-------------------------------------------------------------------------------

\begin{figure*}[!t]
\centering
\begin{subfigure}{0.49\columnwidth}
\includegraphics[width=\columnwidth]{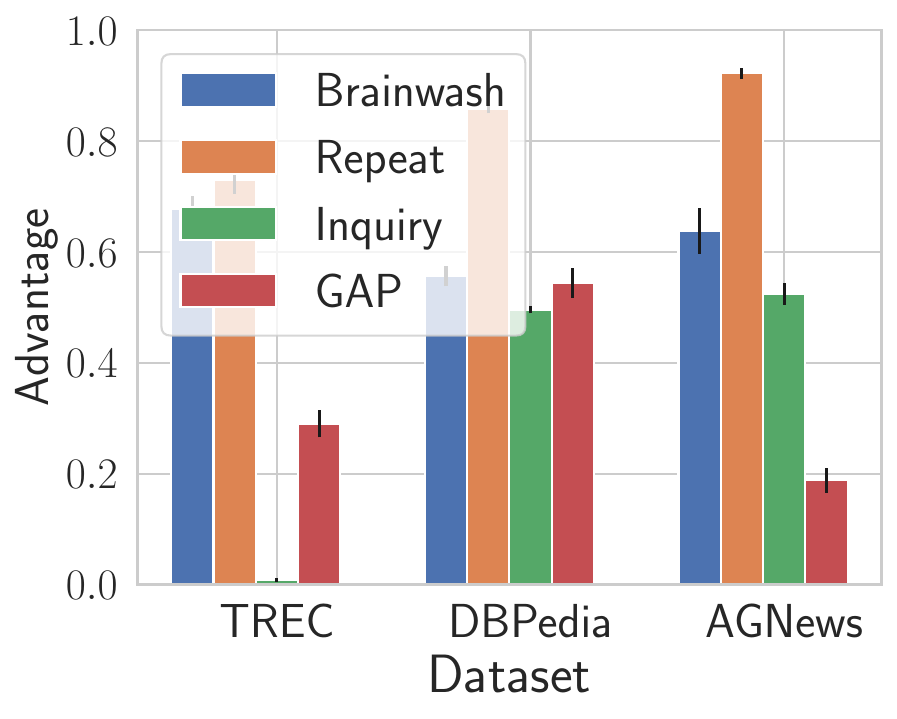}
\caption{GPT2-XL}
\label{fig:compare_one_gpt2xl}
\end{subfigure}
\begin{subfigure}{0.49\columnwidth}
\includegraphics[width=\columnwidth]{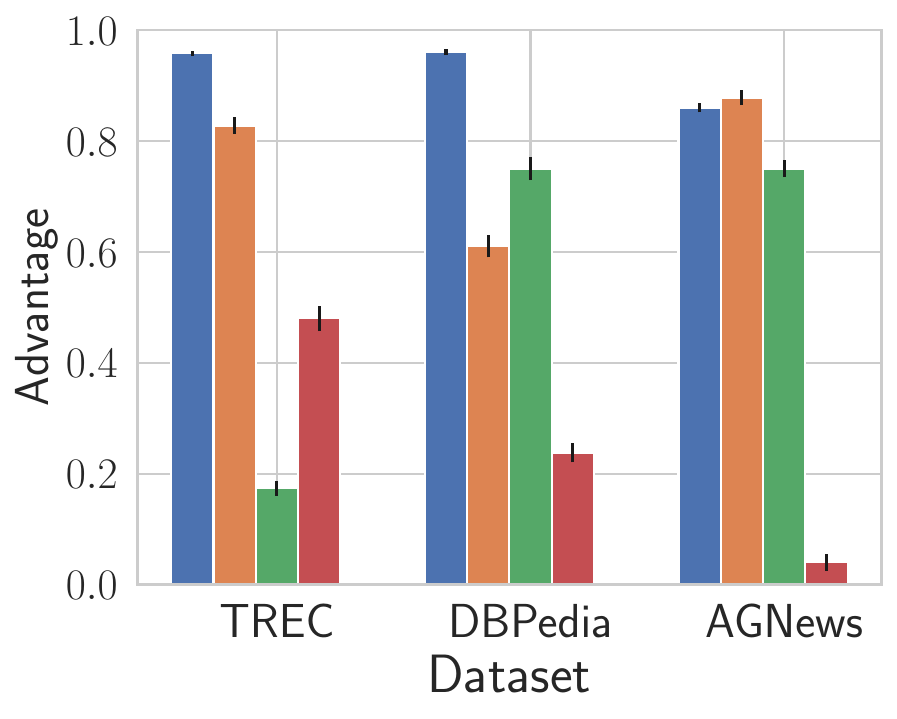}
\caption{LLaMA}
\label{fig:compare_one_llama}
\end{subfigure}
\begin{subfigure}{0.49\columnwidth}
\includegraphics[width=\columnwidth]{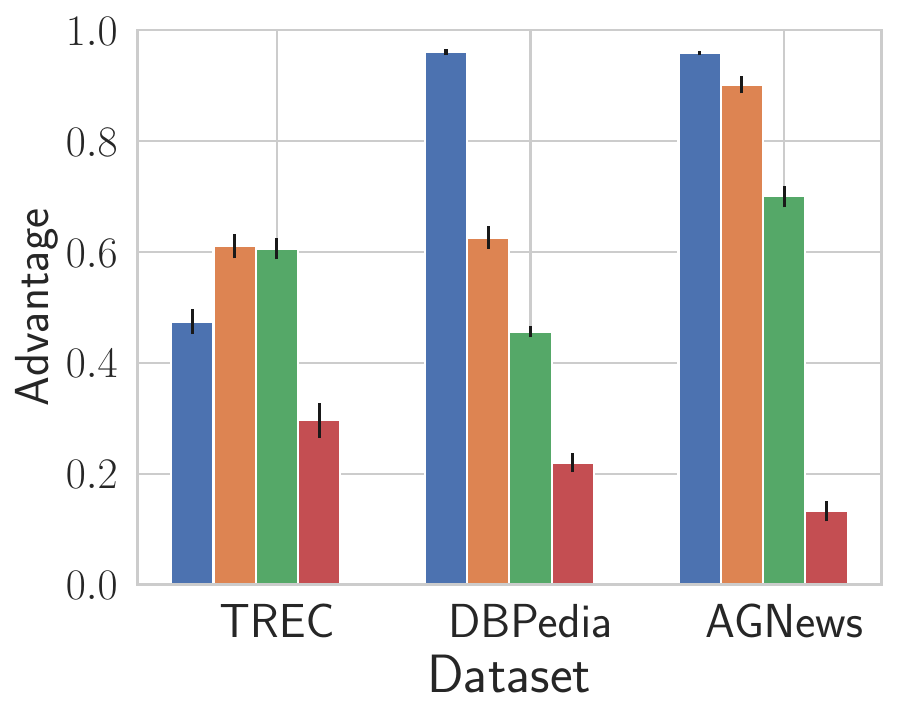}
\caption{Vicuna}
\label{fig:compare_one_vicuna}
\end{subfigure}
\begin{subfigure}{0.49\columnwidth}
\includegraphics[width=\columnwidth]{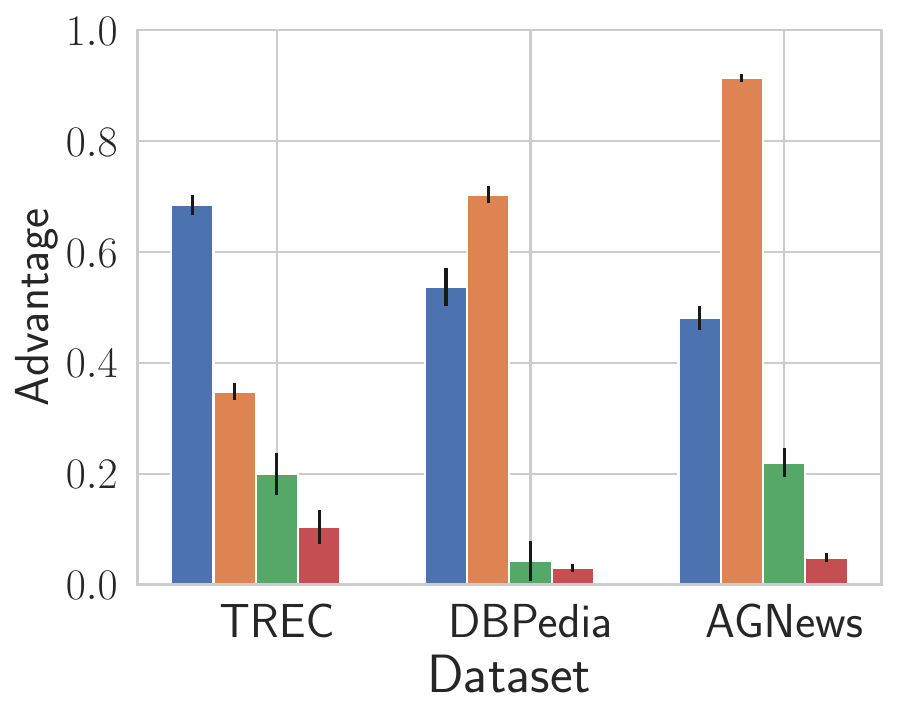}
\caption{GPT-3.5}
\label{fig:compare_one_gpt35}
\end{subfigure}
\caption{Comparison of attack performance across three datasets and four language models, highlighting the consistent efficacy of Brainwash and Repeat attacks, alongside the variable performance of Inquiry and GAP attacks contingent on model architecture.}
\label{figure:compare_one}
\end{figure*}

%-------------------------------------------------------------------------------
\subsection{Experimental Setup}
\label{sec:setup}
%-------------------------------------------------------------------------------

\mypara{Language Models}
We evaluate our attacks on four representative language models, including GPT2-XL~\cite{RWCLAS19}, LLaMA~\cite{TLIMLLRGHARJGL23}, Vicuna~\cite{Vicuna}, and GPT-3.5~\cite{chatgpt_api}.
GPT2-XL is a 1.5B parameter version of GPT-2 developed by OpenAI. 
For LLaMA and Vicuna, we utilize their 7B version and version 1.5 (Vicuna-7b-v1.5), respectively. 
We access GPT-3.5 through its official API with the version name gpt-3.5-turbo-0613, which was released on June 13th. 
We also examine the impact of different versions of this model in ~\autoref{sec:overtime}, providing insights into how variations in model versions affect our attack outcomes.   
This selection spans fully open-source to fully closed commercial models, demonstrating the applicability of our attacks across a wide spectrum.

\mypara{Datasets}
We assess the impact of our attacks on three benchmark text classification datasets: AGNews~\cite{ZZL15}, a 4-class News Topic Classification dataset; TREC~\cite{LR02}, a 6-class Question Classification dataset; and DBPedia~\cite{ZZL15}, a 14-class Ontology Classification dataset.
We structure the prompts according to the template designed by Zhao et al.~\cite{ZWFKS21}, known for its effective performance, with illustrative examples provided in~\refappendix{append:prompt}.
It is worth noting that our objective is to determine if the sample is included in the prompt. 
Therefore, there is no requirement to ensure that these datasets are not utilized in training the pretrained models. 
Given LLMs are trained on extensive datasets, we hypothesize that they do not strongly memorize any specific dataset. 
Therefore, we anticipate minimal impact on performance. 
We further explore the influence of memorization on attack performance in \refappendix{append:memorization_influence}.

\mypara{Evaluation Settings}
The evaluation setting in our work differs from traditional membership inference attack settings, where training datasets typically comprise thousands or tens of thousands of data samples. 
In these cases, adversaries receive both the training dataset and an equivalently sized testing dataset, tasked with determining the membership status for all samples within this mixed dataset. 
However, the context shifts for In-Context Learning, where membership pertains to a smaller subset, typically fewer than eight samples. 
Consequently, evaluating attack performance through a single run may yield unrepresentative results.

Instead of conducting a single experiment, we repeat the experiment 500 times and leverage the average performance as our final result. 
This experimental design, previously employed in studies targeting In-Context Learning~\cite{WWBZS23,DDYPB23}, enhances the robustness and reliability of our assessments.

Each experiment entails the construction of a target prompt based on specified hyperparameters, such as the number of demonstrations. 
Subsequently, we assess the membership status of two target samples: one sample selected from the prompt is labeled as a member, while another is randomly chosen and designated as a nonmember. 

To facilitate this approach, the dataset is initially deduplicated and randomly divided into two parts: the demo part, containing samples utilized for prompt construction, and the test part, housing samples earmarked for testing. 
For each experiment, we randomly select samples from the demo part based on the prompt design to construct the prompt. 
Simultaneously, one sample from the test part is randomly chosen and labeled as a nonmember in the experiment. 
In this paper, we repeat the experiment 500 times, equating to the labeling of a dataset with 500 members and 500 nonmembers.

\mypara{Evaluation Metrics}
We consider two widely applied metrics to evaluate the attack performance:
\begin{itemize}
    \item \textbf{Advantage}~\cite{YGFJ18,SSM19}: This metric, denoted as 
    \[\emph{Adv}=2\times(\emph{Acc}-0.5),\]
    measures the advantage over random guessing, which offers an average-case evaluation of the attack's effectiveness. 
    Following previous work~\cite{YGFJ18,SSM19}, the metric is multiplied by 2 to scale a 100\% accurate attack performance to 1, while random guessing remains at 0. 
    A higher advantage implies better-than-random performance, providing a comprehensive perspective on the attack's overall effectiveness.
    \item \textbf{Log-scale ROC Analysis}~\cite{CCNSTT22}: This metric focuses on the true-positive rate at low false-positive rates, effectively capturing the worst-case attack performance.
\end{itemize}

Given that determining membership status is a binary classification task, with an equal number of positive and negative samples, the advantage metric provides an intuitive measure of performance. 
An advantage of 0 signifies random guessing, while an advantage of 1 indicates an accurate prediction of all membership statuses. 
This simplicity aids in the straightforward interpretation of the performance of the attacks.

For all proposed attacks, we employ the advantage metric to gauge average-case performance. 
In the case of the Repeat and Brainwash attacks, we additionally utilize the log-scale ROC curve to depict their worst-case performance. 
It's important to note that we cannot present worst-case performance for all attacks since, in the other two, only hard membership predictions are obtainable.

\begin{figure*}[!t]
\centering
\begin{subfigure}{0.49\columnwidth}
\includegraphics[width=\columnwidth]{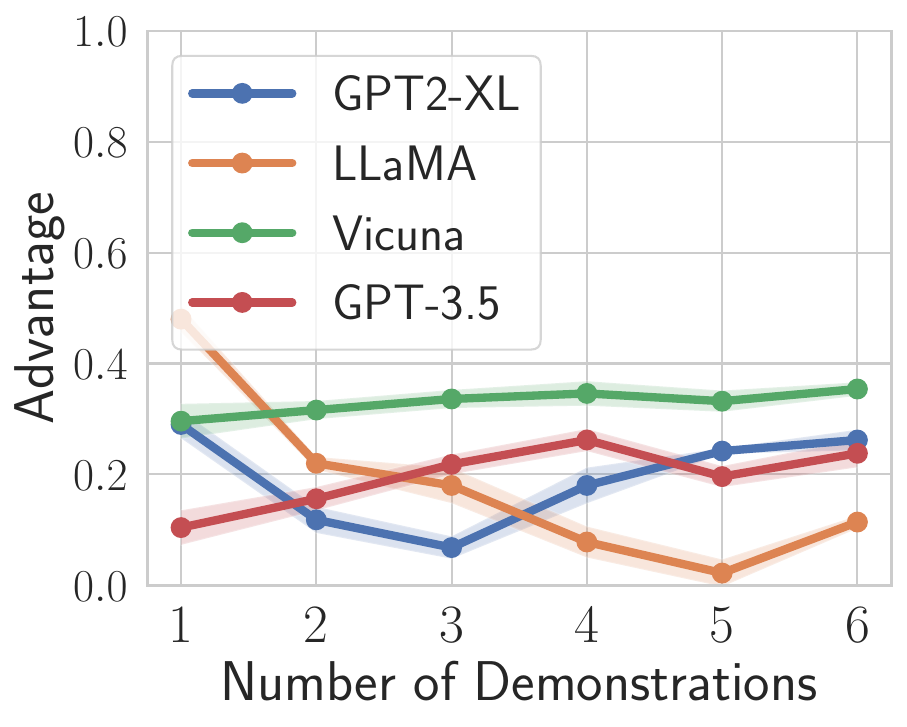}
\caption{GAP}
\label{fig:demonumber_gap_trec}
\end{subfigure}
\begin{subfigure}{0.49\columnwidth}
\includegraphics[width=\columnwidth]{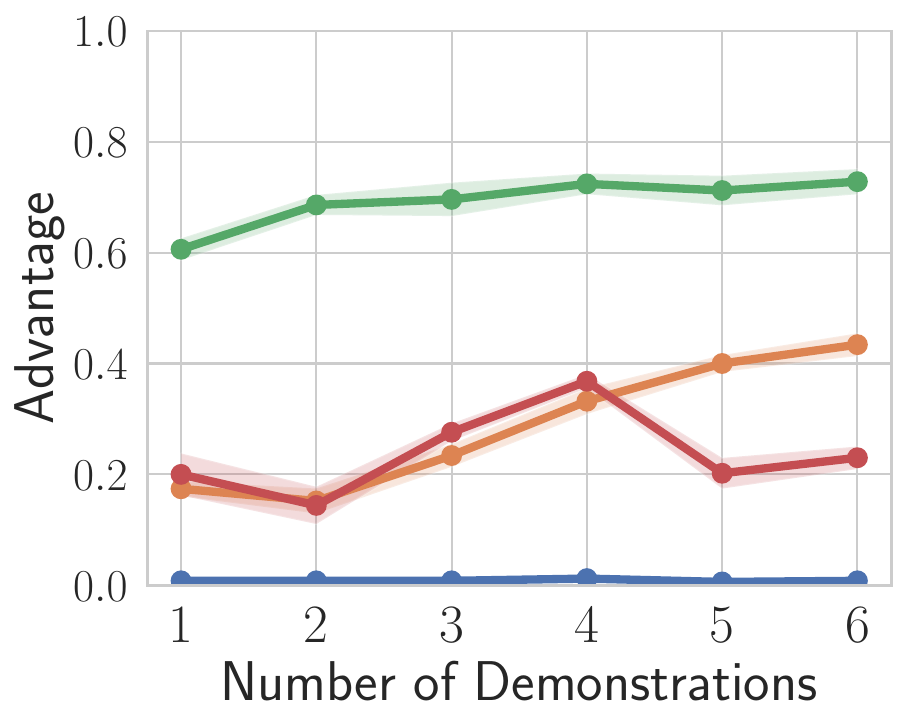}
\caption{Inquiry}
\label{fig:demonumber_inquiry_trec}
\end{subfigure}
\begin{subfigure}{0.49\columnwidth}
\includegraphics[width=\columnwidth]{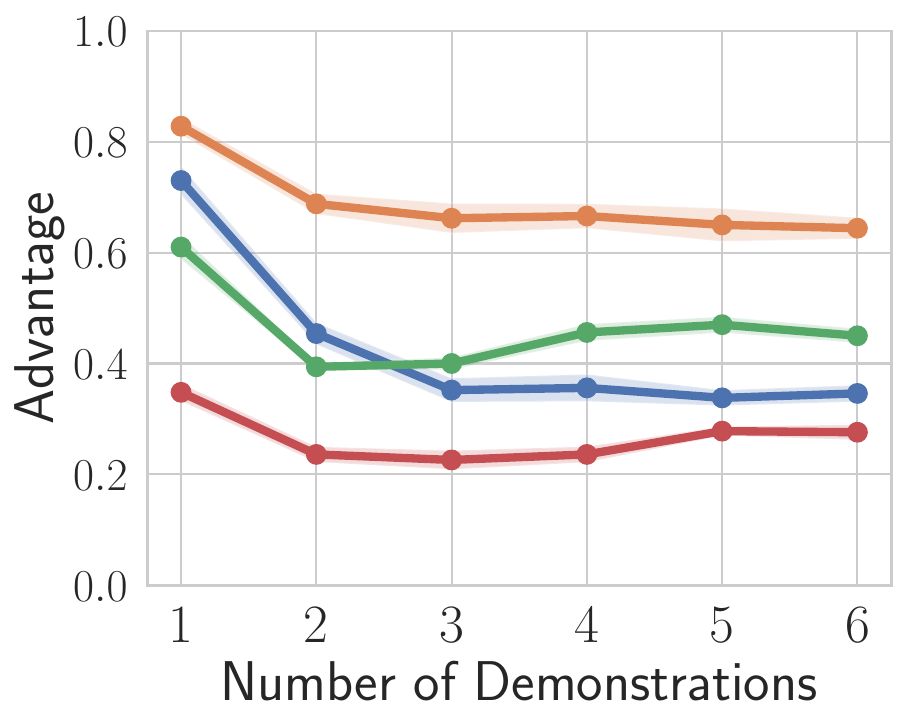}
\caption{Repeat}
\label{fig:demonumber_repeat_trec}
\end{subfigure}
\begin{subfigure}{0.49\columnwidth}
\includegraphics[width=\columnwidth]{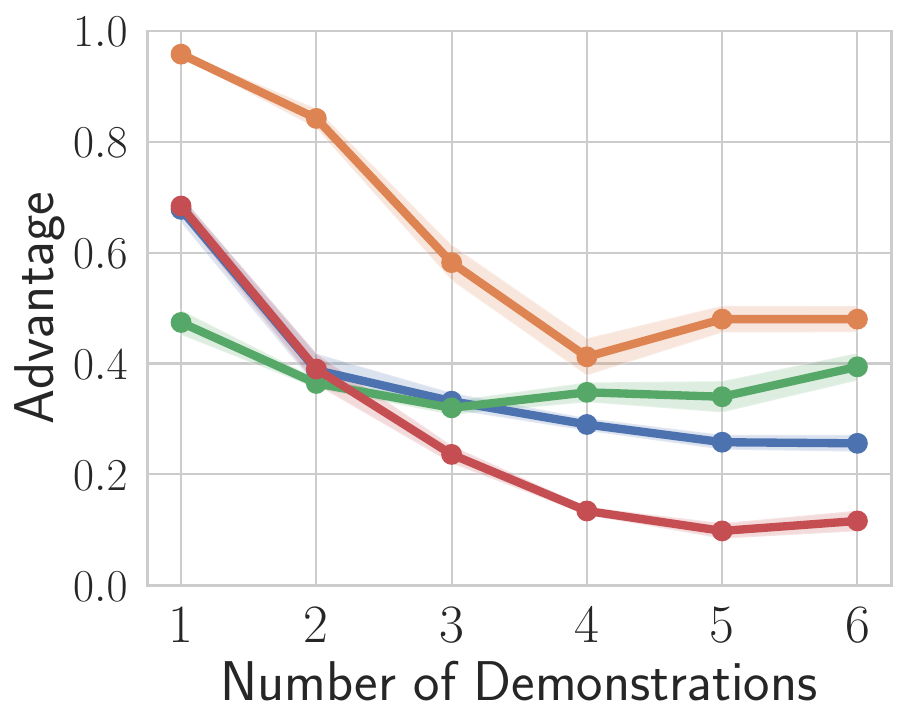}
\caption{Brainwash}
\label{fig:demonumber_brainwash_trec}
\end{subfigure}
\caption{Membership Inference Attack (MIA) performance with varying numbers of demonstrations in the prompt, illustrating the influence of demonstration quantity on the efficacy of Repeat and Brainwash attacks. 
Our experiments are conducted on the TREC dataset.}
\label{figure:demonumber_trec}
\end{figure*}

\begin{figure}[!t]
\centering
\begin{subfigure}{0.49\columnwidth}
\includegraphics[width=\columnwidth]{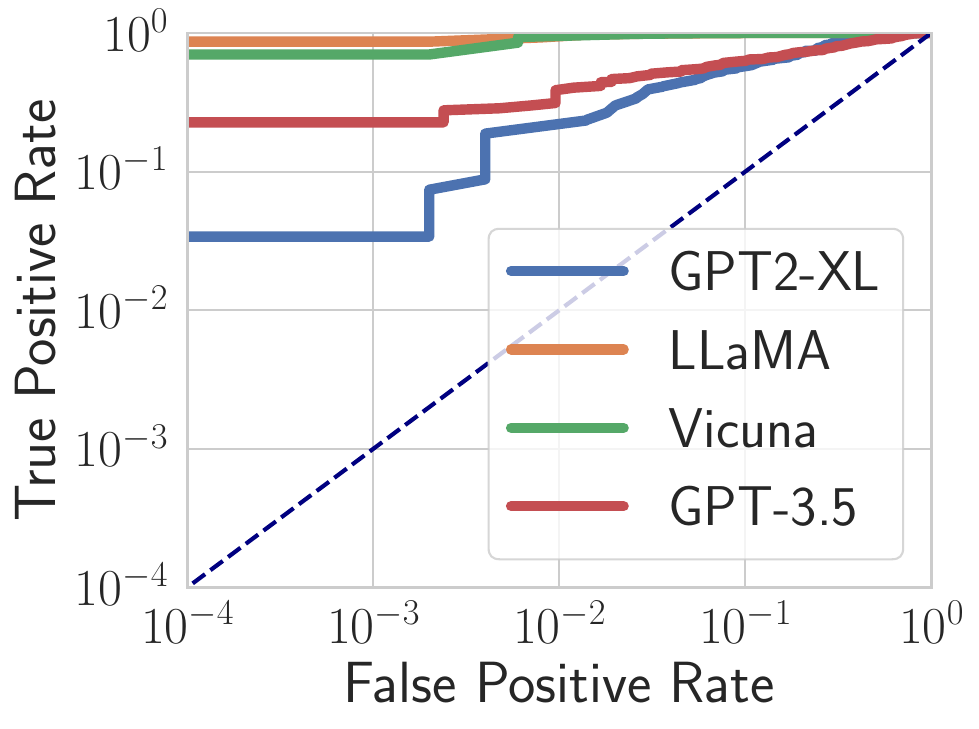}
\caption{Brainwash}
\label{fig:roc_one_brainwash}
\end{subfigure}
\begin{subfigure}{0.49\columnwidth}
\includegraphics[width=\columnwidth]{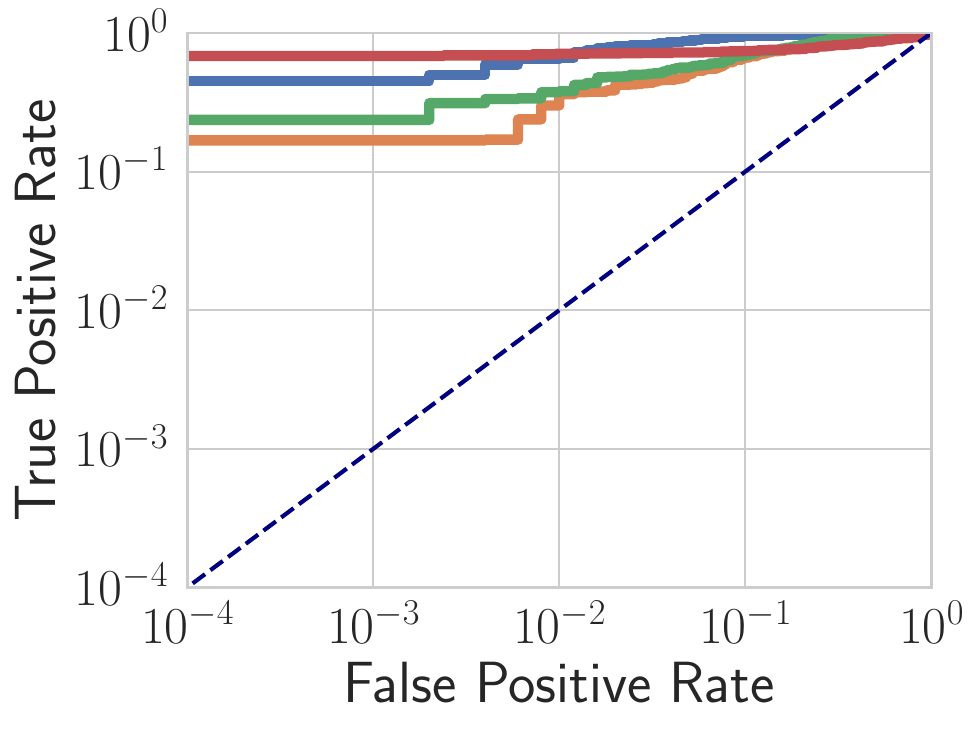}
\caption{Repeat}
\label{fig:roc_one_repeat}
\end{subfigure}
\caption{Log-scale Receiver Operating Characteristic (ROC) curve depicting the worst-case performance of Brainwash and Repeat attacks, revealing their efficacy in discerning ambiguous samples.}
\label{figure:roc_one}
\end{figure}

%-------------------------------------------------------------------------------
\subsection{Results}
%-------------------------------------------------------------------------------

We start by evaluating the performance of our attacks under the basic setting, where the $\prompt$ contains only one demonstration: $\prompt = \{\instruct, s(x_1, y_1)\}$.
Here, the adversary aims to determine whether the target sample $x$ is contained in $\prompt$, i.e., $x=x_1$ or $x \neq x_1$.

We report the advantage of all four attacks in~\autoref{figure:compare_one}. 
As we can see, Brainwash and Repeat attacks consistently exhibit strong performance across all four language models. 
This remarkable performance is particularly evident with LLaMA and Vicuna, as demonstrated in~\autoref{fig:compare_one_llama} and~\autoref{fig:compare_one_vicuna}. 
For example, Brainwash achieves nearly 100\% advantage on 5 out of 6 tasks with LLaMA and Vicuna.

In contrast, the performance of Inquiry and GAP attacks varies significantly depending on the model architecture. 
For GPT2-XL, GAP attack even achieves 54.4\% advantage on DBPedia, although this is the optimal performance it can achieve across all datasets and models. 
Furthermore, with LLaMA, the Inquiry attack achieves 75.0\% advantage on AGNews, showcasing strong performance even with relatively straightforward approaches. 
However, for GPT-3.5, both attacks prove ineffective, with a performance close to random guessing. 
Encouragingly, Brainwash and Repeat attacks, while showing a slight performance decrease, maintain effectiveness in inferring membership status on GPT-3.5.
This observation highlights the prevalence of membership leakage vulnerabilities in large language models, even when the model solely outputs text information.

Notably, while the Brainwash and Repeat attacks consistently outperform the other two, they show varying advantages under different model architectures. 
As \autoref{figure:compare_one} shows, for the GPT family of models (GPT2-XL and GPT-3.5), the Repeat attack outperforms Brainwash in most cases, suggesting that the generative behaviors of these models on members and non-members are quite different when queries starting from a few words are given complementary words to finish with.
However, for LLaMA and Vicuna, they show greater vulnerability to the Brainwash attack, suggesting that these models are more firm or believe in the knowledge they have gained from $\prompt$.
The question of why these large language models have different properties and behaviors is beyond the scope of this work, and we leave it to more relevant research areas.

We further report the worst-case performance (log-scale ROC) of the Brainwash and Repeat attacks in~\autoref{figure:roc_one}. 
We can observe that both Brainwash and Repeat attacks exhibit remarkable performance, particularly in the low false positive area. 
This observation implies the effectiveness of our attacks in determining the membership status of samples that are hard to differentiate.

%-------------------------------------------------------------------------------
\subsection{Influence of Number of Demonstrations}
%-------------------------------------------------------------------------------

In the previous section, we established the effectiveness of our attacks under the scenario where the $\prompt$ solely comprises one demonstration. 
However, in a more practical scenario, language model owners often leverage multiple demonstrations to construct $\prompt$ to enhance performance.
In this section, we explore the performance of our proposed attack under the influence of the number of demonstrations.

It is noteworthy that an increased number of demonstrations does not unilaterally enhance performance, as an extended prompt incurs higher costs in terms of tokens and may be constrained by input limitations of the language model. 
Therefore, in this section, we vary the number of demonstrations from 1 to 6 to assess its impact on attack performance. 
We limited the number of demonstrations to 6 due to the input size restrictions of GPT2-XL, and further evaluated the long context performance in~\refappendix{append:longcontext}.

\begin{figure}[!t]
\centering
\begin{subfigure}{0.49\columnwidth}
\includegraphics[width=\columnwidth]{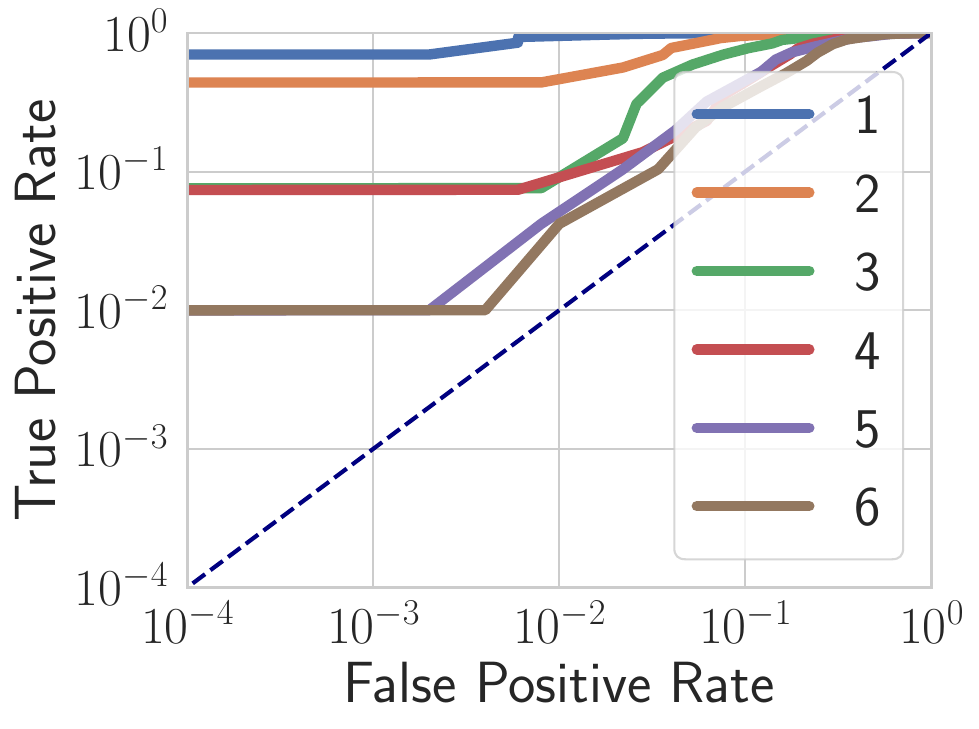}
\caption{Brainwash}
\label{fig:roc_shots_brainwash}
\end{subfigure}
\begin{subfigure}{0.49\columnwidth}
\includegraphics[width=\columnwidth]{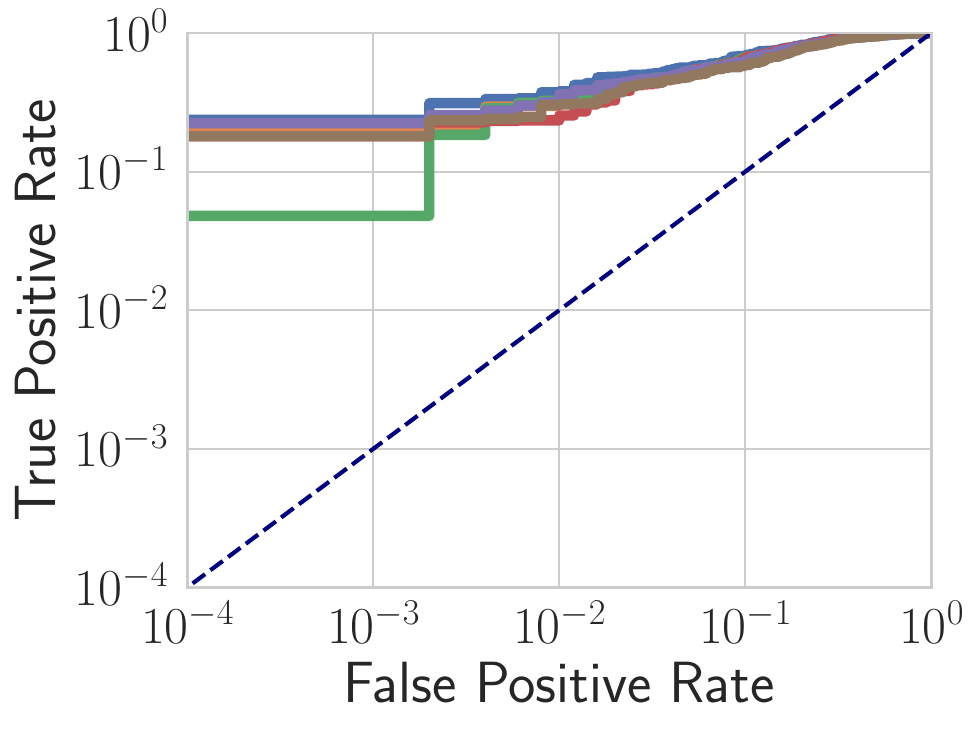}
\caption{Repeat}
\label{fig:roc_shots_repeat}
\end{subfigure}
\caption{Log-scale ROC curve illustrating the worst-case performance of Membership Inference Attacks with varying numbers of demonstrations in the prompt, emphasizing the consistent superiority of one demonstration over multiple demonstrations.}
\label{figure:roc_shots}
\end{figure}

\begin{figure*}[!t]
\centering
\begin{subfigure}{0.49\columnwidth}
\includegraphics[width=\columnwidth]{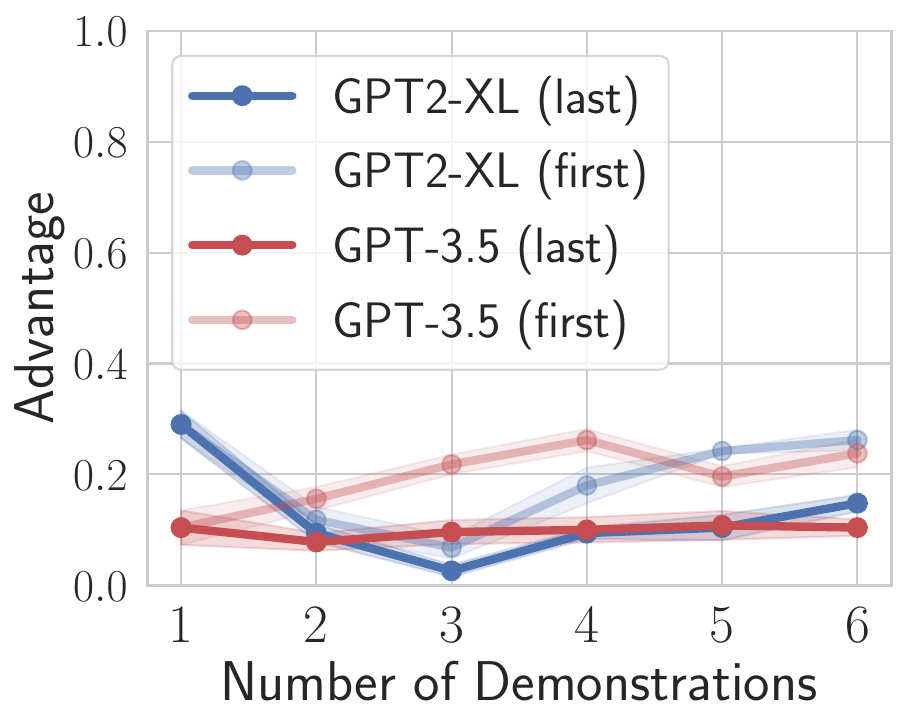}
\caption{GAP}
\label{fig:demonumber_pos_comp_gap_trec}
\end{subfigure}
\begin{subfigure}{0.49\columnwidth}
\includegraphics[width=\columnwidth]{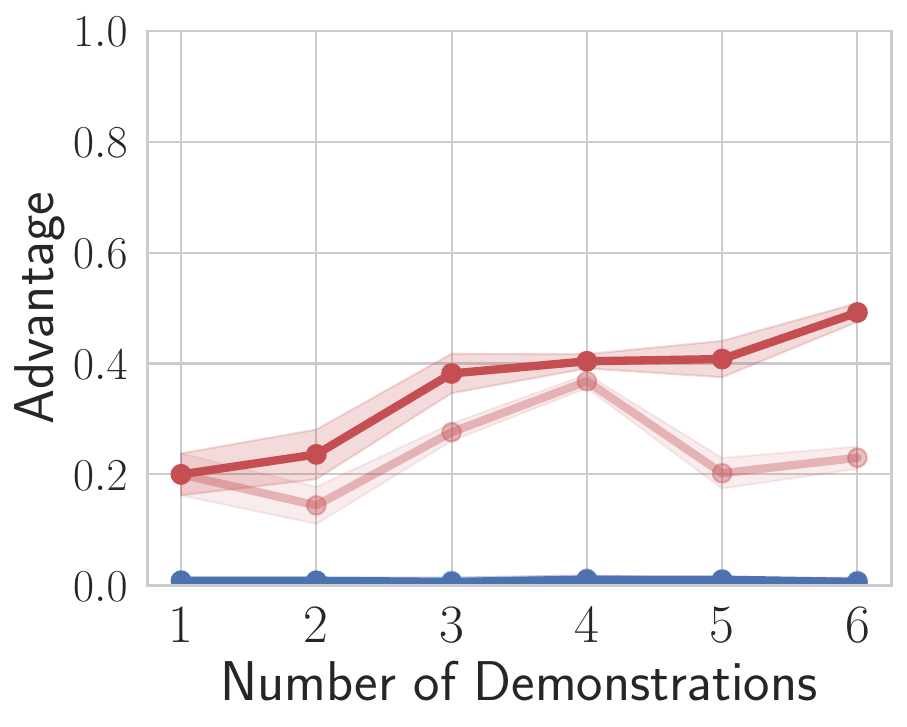}
\caption{Inquiry}
\label{fig:demonumber_pos_comp_inquiry_trec}
\end{subfigure}
\begin{subfigure}{0.49\columnwidth}
\includegraphics[width=\columnwidth]{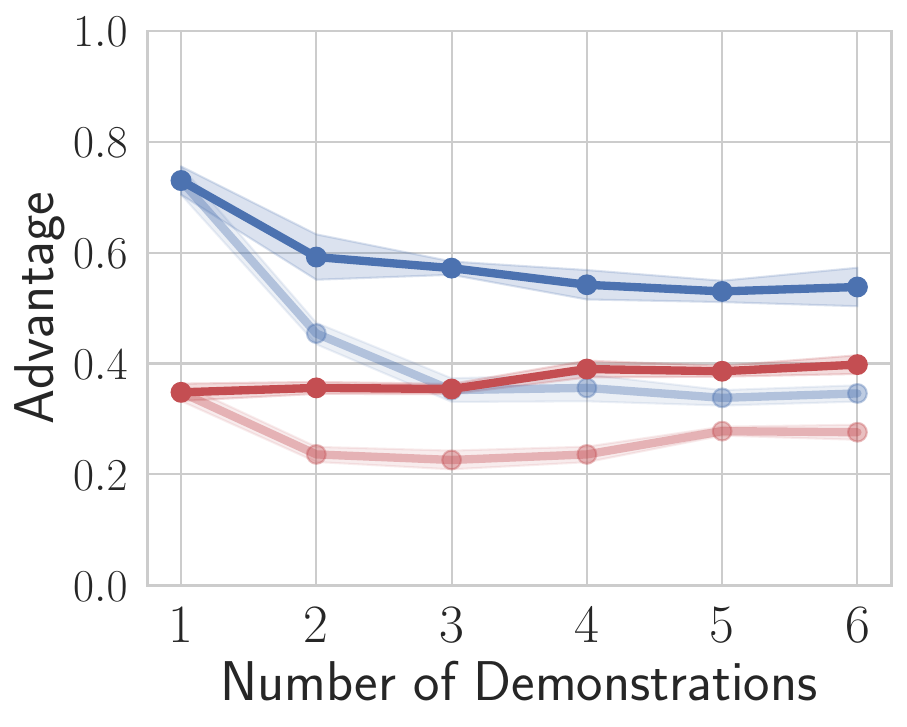}
\caption{Repeat}
\label{fig:demonumber_pos_comp_repeat_trec}
\end{subfigure}
\begin{subfigure}{0.49\columnwidth}
\includegraphics[width=\columnwidth]{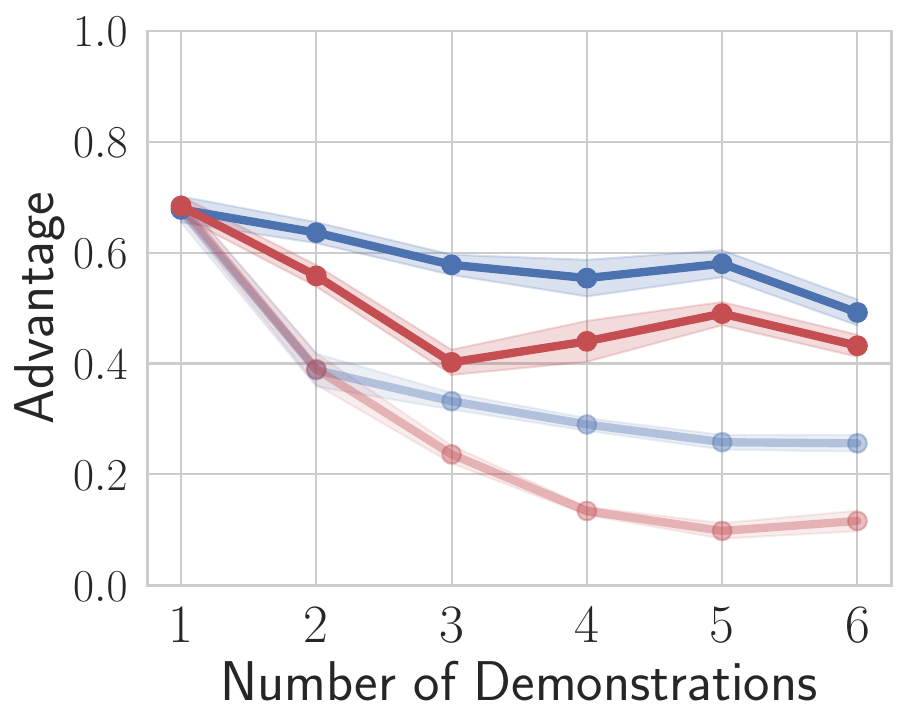}
\caption{Brainwash}
\label{fig:demonumber_pos_comp_brainwash_trec}
\end{subfigure}
\caption{Comparative analysis of Membership Inference Attack (MIA) performance targeting the first and last demonstration, underscores the impact of the distance between the target sample and the query on model memorization. 
Our experiments are conducted on the TREC dataset.}
\label{figure:demonumber_pos_comp_trec}
\end{figure*}

\begin{figure*}[!t]
\centering
\begin{subfigure}{0.49\columnwidth}
\includegraphics[width=\columnwidth]{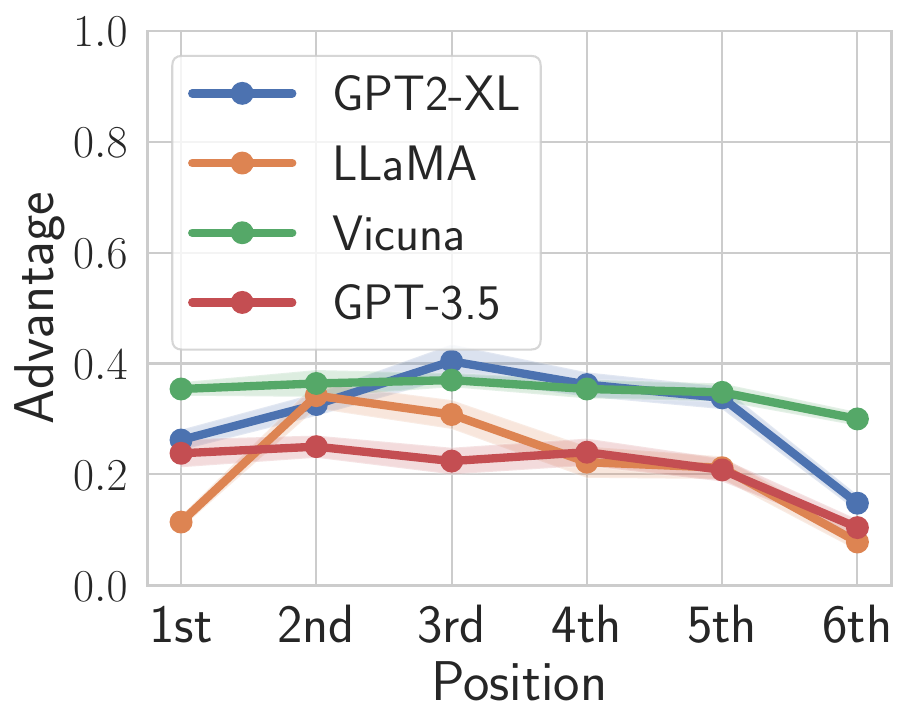}
\caption{GAP}
\label{fig:position_gap_trec}
\end{subfigure}
\begin{subfigure}{0.49\columnwidth}
\includegraphics[width=\columnwidth]{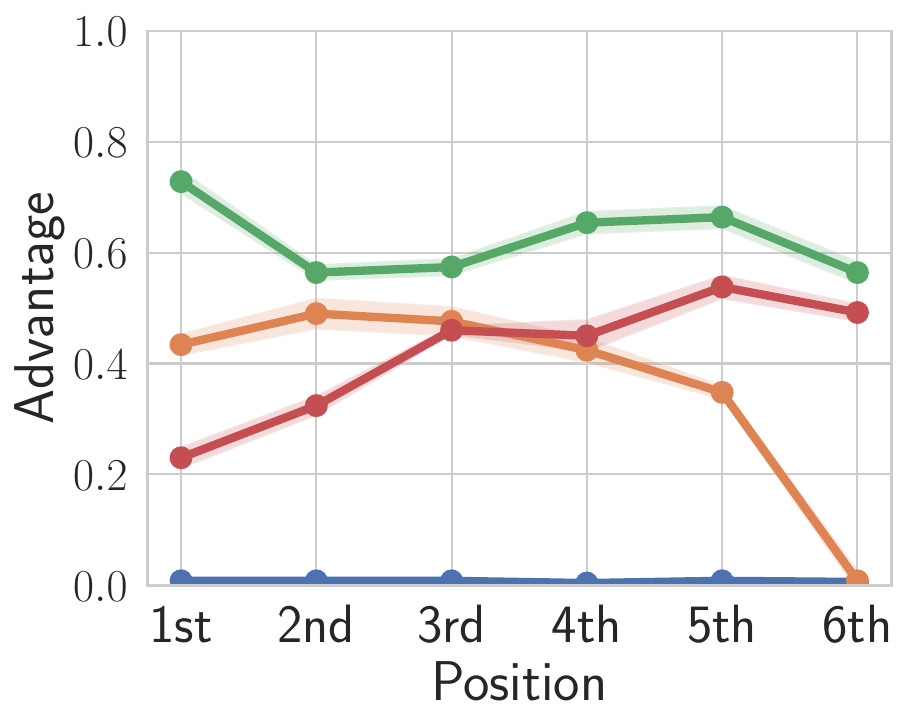}
\caption{Inquiry}
\label{fig:position_inquiry_trec}
\end{subfigure}
\begin{subfigure}{0.49\columnwidth}
\includegraphics[width=\columnwidth]{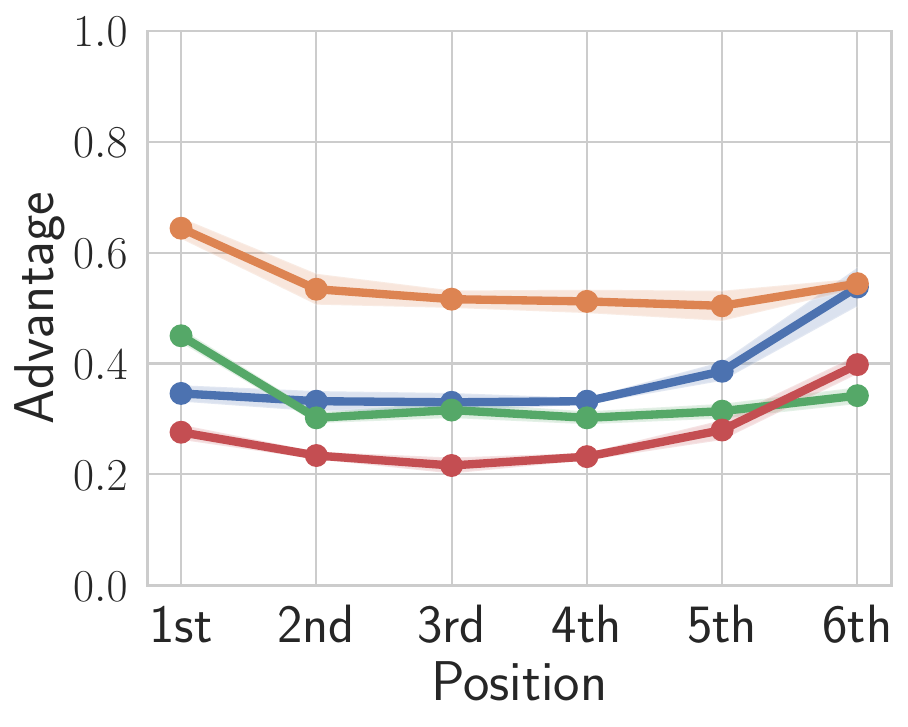}
\caption{Repeat}
\label{fig:position_repeat_trec}
\end{subfigure}
\begin{subfigure}{0.49\columnwidth}
\includegraphics[width=\columnwidth]{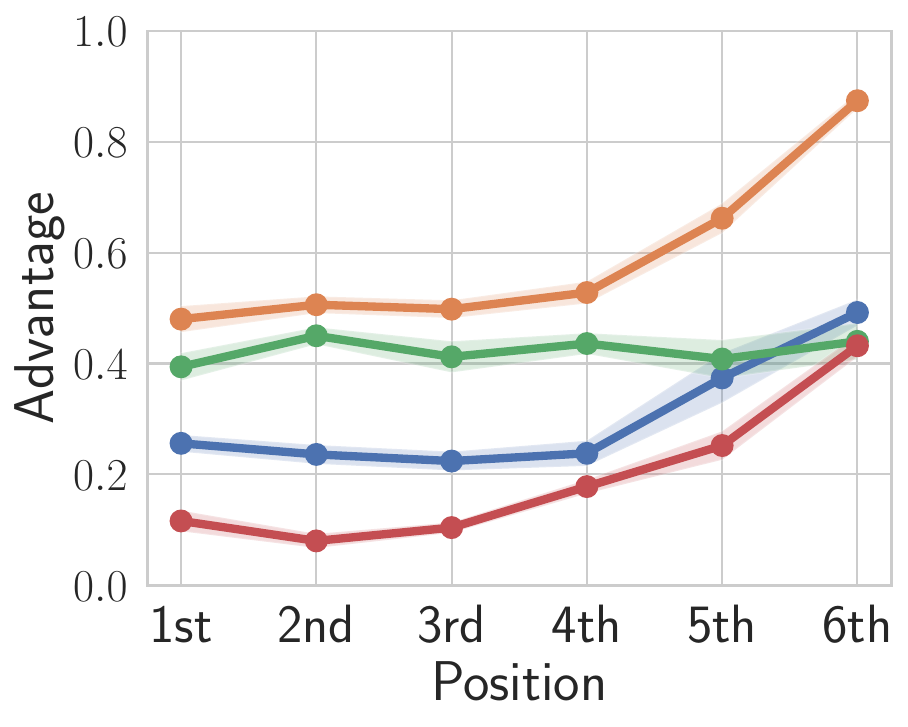}
\caption{Brainwash}
\label{fig:position_brainwash_trec}
\end{subfigure}
\caption{Exploration of attack performance across different positions (1st to 6th) of demonstrations within a prompt.
Results reveal varying vulnerabilities for Repeat and Brainwash attacks.
Notably, demonstrations in the middle exhibit a lower vulnerability compared to those positioned at the beginning or end. 
Our experiments are conducted on the TREC dataset.}
\label{figure:position_trec}
\end{figure*}

Moreover, we consider the positional influence of demonstrations within the prompt. 
Specifically, we posit that for $\prompt = \{\instruct, s(x_1, y_1)\dots s(x_k, y_k)\}$ containing $k$ demonstrations, the impact of $x_1$ and $x_k$ should be different.
The reason for this argument is that the demonstrations are entered into the language model sequentially, and the model's memory for the first demonstration should be significantly influenced by subsequent demonstrations compared to the last demonstration. 
Consequently, in the ensuing experiments probing the influence of the demonstration number ($k$), we examine the effects on the first and last demonstrations (i.e., $x_1$ and $x_k$) separately.

We first report the effect of $k$ on inferring the first demonstration in \autoref{figure:demonumber_trec}.
We can see that while the GAP and Inquiry attacks are insensitive to the number of demonstrations, the remaining two attacks show a clear trend between it and attack performance.
Specifically, both the Repeat and Brainwash attacks obtain optimal results when there is only one demonstration, and the attack performance gradually decreases as the number of demonstrations increases.
We further report the worst-case performance in ~\autoref{figure:roc_shots}.
We can find that attacks against one demonstration consistently outperform scenarios with more demonstrations, even in terms of low false positives.

We then report the effect of $k$ on inferring the last demonstration.
For comparison purposes, we present the results for both the first and the last presentation.
In addition, in order to observe the trend more clearly, we only present the results in two language models: GPT2-XL and GPT-3.5.
First, as shown in \autoref{figure:demonumber_pos_comp_trec}, we can get a similar observation that there is no significant trend for GAP and Inquiry attacks, but a decreasing trend for Repeat and Brainwash attacks.
Therefore, we next discuss the findings based on the latter two powerful attacks (\autoref{fig:demonumber_pos_comp_repeat_trec} and \autoref{fig:demonumber_pos_comp_brainwash_trec}).
These observations suggest that attack performance is negatively correlated with the number of demonstrations, regardless of which demonstrations an adversary aims at to infer membership.
These observations also suggest that more demonstrations will not only improve model performance, but also reduce membership leakage.
Furthermore, we can find that the attack performance of the first demonstration is more susceptible to the number of demonstrations compared to the last demonstration, as shown by the steeper trend of the first demonstration.
This observation validates our previously mentioned argument that the model's memory on the first demonstration is more likely to be affected than on the last demonstration, as the next few demonstrations may overwrite the knowledge from the first demonstration.

\begin{figure}[!t]
\centering
\begin{subfigure}{0.49\columnwidth}
\includegraphics[width=\columnwidth]{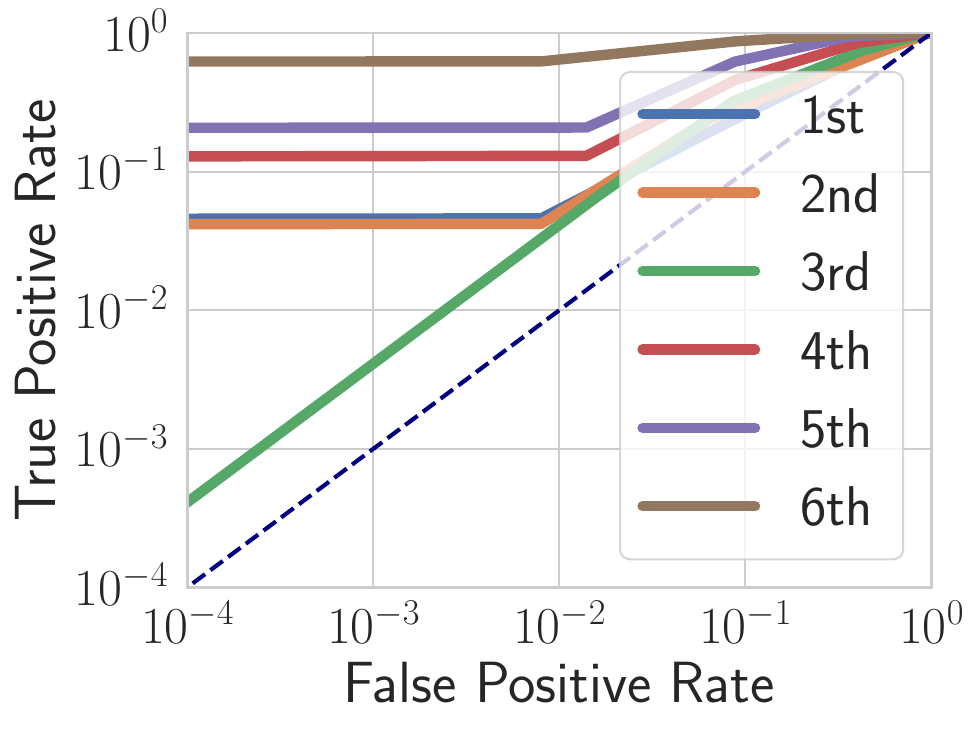}
\caption{Brainwash}
\label{fig:roc_position_brainwash}
\end{subfigure}
\begin{subfigure}{0.49\columnwidth}
\includegraphics[width=\columnwidth]{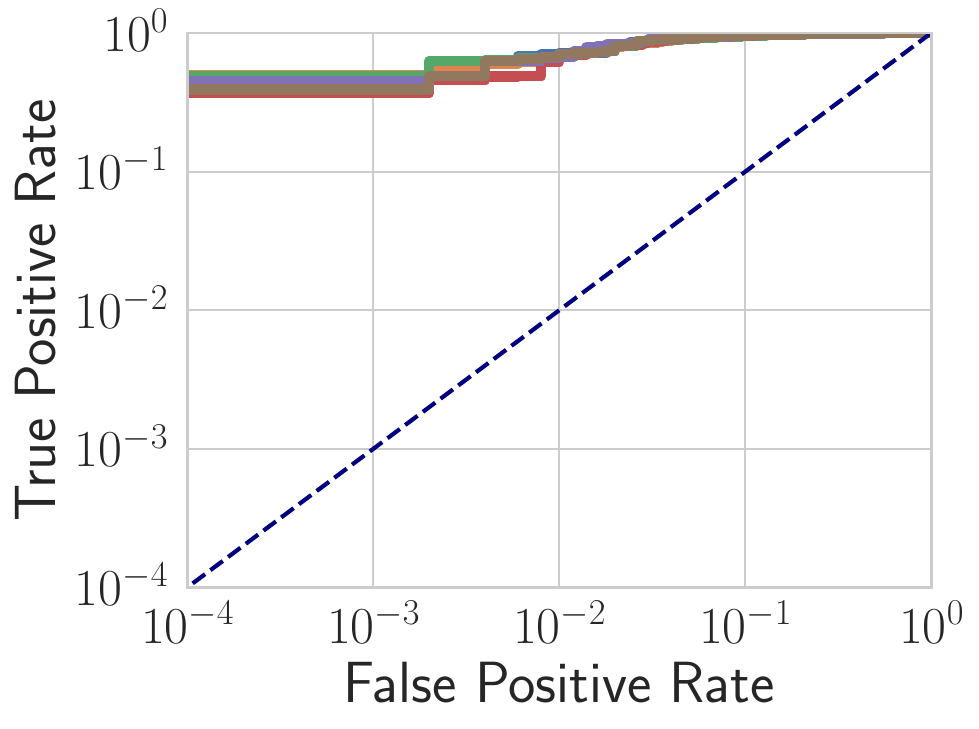}
\caption{Repeat}
\label{fig:roc_position_repeat}
\end{subfigure}
\caption{Log-scale ROC curve confirms that demonstrations in the middle exhibit reduced vulnerability compared to those located at the beginning or end, even in the worst-case scenario. 
We use LLaMA as an example here.}
\label{figure:roc_position}
\end{figure}

\begin{figure*}[!t]
\centering
\begin{subfigure}{0.6\columnwidth}
\includegraphics[width=\columnwidth]{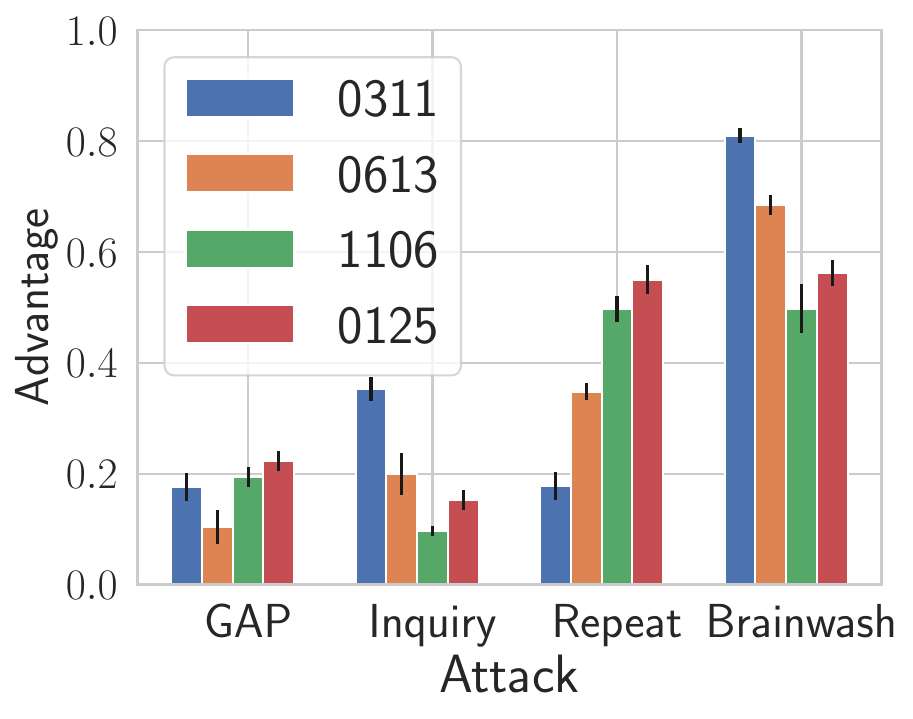}
\caption{One demonstration}
\label{fig:overtime_1_end}
\end{subfigure}
\begin{subfigure}{0.6\columnwidth}
\includegraphics[width=\columnwidth]{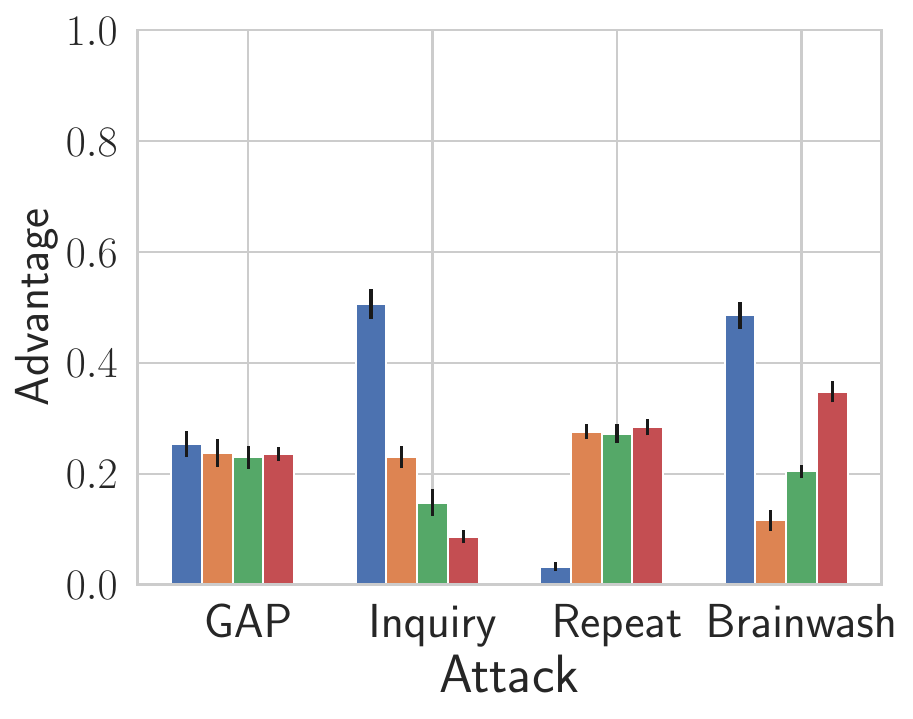}
\caption{Six demonstrations (first)}
\label{fig:overtime_6_begin}
\end{subfigure}
\begin{subfigure}{0.6\columnwidth}
\includegraphics[width=\columnwidth]{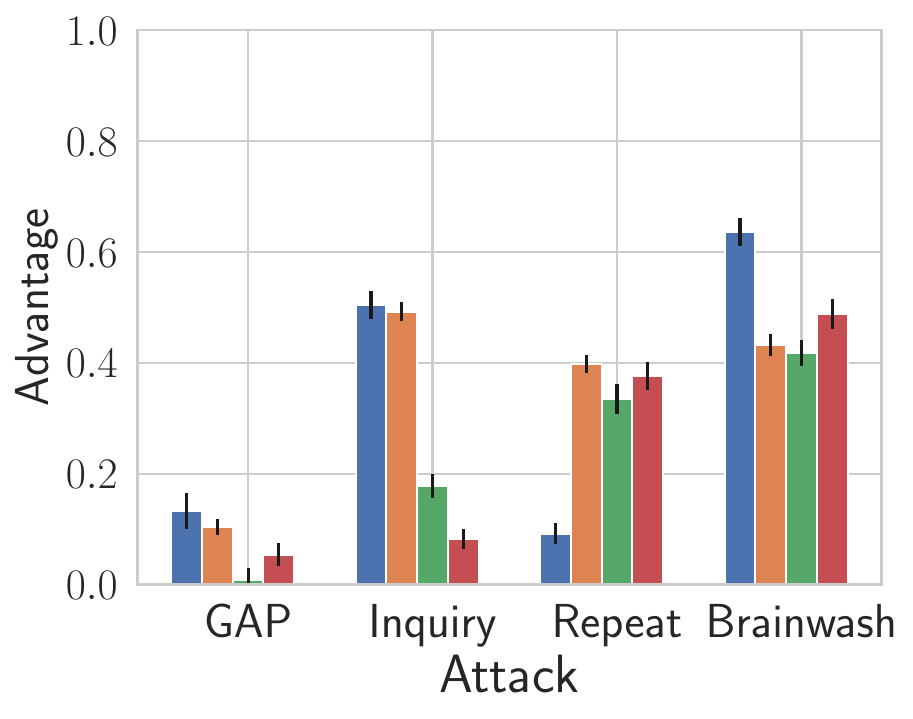}
\caption{Six demonstrations (last)}
\label{fig:overtime_6_end}
\end{subfigure}
\caption{Evolution of Attack Performance: Comparative analysis of attack performance on four GPT-3.5 API versions (gpt-3.5-turbo-0301, gpt-3.5-turbo-0613, gpt-3.5-turbo-1106, and gpt-3.5-turbo-0124) over a ten-month period. 
Results demonstrate that the robustness of commercial models like GPT-3.5 doesn’t monotonically increase over time. 
We conduct our experiments on the TREC dataset. 
Experiments on the DBPedia dataset derive the same conclusion, detailed results can be found in~\refappendix{append:overtime_dbpedia}.}
\label{figure:overtime}
\end{figure*}

%-------------------------------------------------------------------------------
\subsection{Influence of the Demonstration Position} 
%-------------------------------------------------------------------------------

The above results suggest that the number of demonstrations has a different effect on the first and last demonstrations.
We next explore in more depth the effect of demonstration position on attack performance.
Concretely, we maintain a consistent number of demonstrations at 6 and use our proposed attacks to infer the membership status of each demonstration, ranging from the $1st$ to the $6th$ position. 

We report the relationship between demonstration position and attack performance in \autoref{figure:position_trec}.
We can observe that demonstrations at different positions exhibit varying vulnerability, particularly evident for our two powerful attacks, Repeat and Brainwash. 
Interestingly, the results show that the worst attack performance is not against the first demonstration.
Instead, demonstrations positioned in the middle sometimes exhibit poor attack performance. 
For instance, as illustrated in~\autoref{fig:position_repeat_trec}, inferring the membership status of the first demonstration in GPT-3.5 yields 27.6\% advantage, while inferring the third demonstration results in only 21.6\% advantage. 
This observation is further validated in the worst-case performance shown in~\autoref{figure:roc_position}. 
Notably, for the Brainwash attack (\autoref{fig:roc_position_brainwash}), the last demonstration attains the best performance, as expected, with the first demonstration ranking second. 
Conversely, the third demonstration (represented by the green line) exhibits the poorest performance.

We emphasize that this finding aligns with previous research~\cite{LLHPBPL23}, which indicates that when large language models encounter long input, they are more prone to focusing on the initial and concluding parts of the context while neglecting the middle section. 
This conclusion may motivate users to strategically place their valuable demonstrations in the middle of the prompt. 
However, this strategic placement may potentially compromise utility. 
Consequently, designing a privacy-preserving positional strategy that balances privacy and utility presents an intriguing problem for further exploration.

%-------------------------------------------------------------------------------
\subsection{Attack Performance Over Time}
\label{sec:overtime}
%-------------------------------------------------------------------------------

In response to the increasing security risks associated with Large Language Models, both researchers and companies are focusing on developing responsible and resilient models capable of withstanding potential threats, including jailbreak attacks. 
A significant area of exploration involves examining how different versions of Large Language Models address these security challenges, particularly concerning text-only membership inference attacks, as proposed in our study.

In this section, we conduct a case study using one of the most influential LLMs, GPT-3.5, to discern how the performance of attacks varies across different versions. 
Since the beginning of 2023, the OpenAI has released four API versions: gpt-3.5-turbo-0301, gpt-3.5-turbo-0613, gpt-3.5-turbo-1106, and the latest gpt-3.5-turbo-0125, with the numerical suffix denoting the release date (the first three in 2023 and the latest in 2024). 
Employing these versions, we execute our proposed four attacks and assess how their performance changes over an eleven-month period.

We first evaluate attack performance with the basic setup of $\prompt$ containing only one demonstration (sampled from the TREC dataset) and report the results in \autoref{fig:overtime_1_end}.
We can find a clear trend in the different patterns exhibited by different versions of GPT-3.5 under attacks.
Notably, the attack performance of the newly released APIs decreases under our Brainwash and Inquiry attacks but increases in the latest version.
In contrast, the Repeat attack demonstrates higher performance on recently released APIs. 
In the case of the GAP attack, which considers the generalization gap in training and testing datasets, the attack differences across the four versions of LLMs are negligible.

Extending our observations to $\prompt$ with multiple demonstrations, as depicted in~\autoref{fig:overtime_6_begin}, where $\prompt$  contains 6 demonstrations, and the target demonstration is positioned at the beginning, reveals an intriguing pattern. 
The older version (gpt-3.5-turbo-0301) exhibits higher attack performance for Brainwash and Inquiry attacks, while Repeat shows significantly lower performance. 
This observation remains consistent regardless of the target demonstration's position, as illustrated in~\autoref{fig:overtime_6_end}, where placing the target demonstration at the end yields conclusions analogous to those when positioned at the beginning.

This observation may stem from synergies between attacks, as observed in previous studies~\cite{SSM19,LWHSZBCFZ22}, different attacks exhibit correlations. 
For instance, defending against adversarial examples could inevitably increase vulnerability to membership inference attacks. 
Given the diverse attack surface against LLMs, protecting against all potential threats becomes exceedingly challenging.

In summary, different versions of GPT-3.5 APIs manifest varied behaviors when subjected to attacks, and no single version outperforms all four types of attacks. 
This understanding helps us navigate the challenge of securing language models in the face of evolving security threats.

%-------------------------------------------------------------------------------
\section{Hybrid Attack}
%-------------------------------------------------------------------------------

Given the observed varieties in the effectiveness of different attacks across models, datasets, and model versions, it becomes a fascinating challenge from an adversary's point of view to devise a combined approach that consistently performs well in different scenarios.
In this section, we present a hybrid attack that combines the strengths of both the Brainwash attack and the Repeat attack to achieve strong performance in all scenarios. 
We chose these two attacks because they are resilient across different models and datasets.

%-------------------------------------------------------------------------------
\subsection{Methodology}
%-------------------------------------------------------------------------------

\begin{figure}[!t]
\centering
\begin{subfigure}{0.49\columnwidth}
\includegraphics[width=\columnwidth]{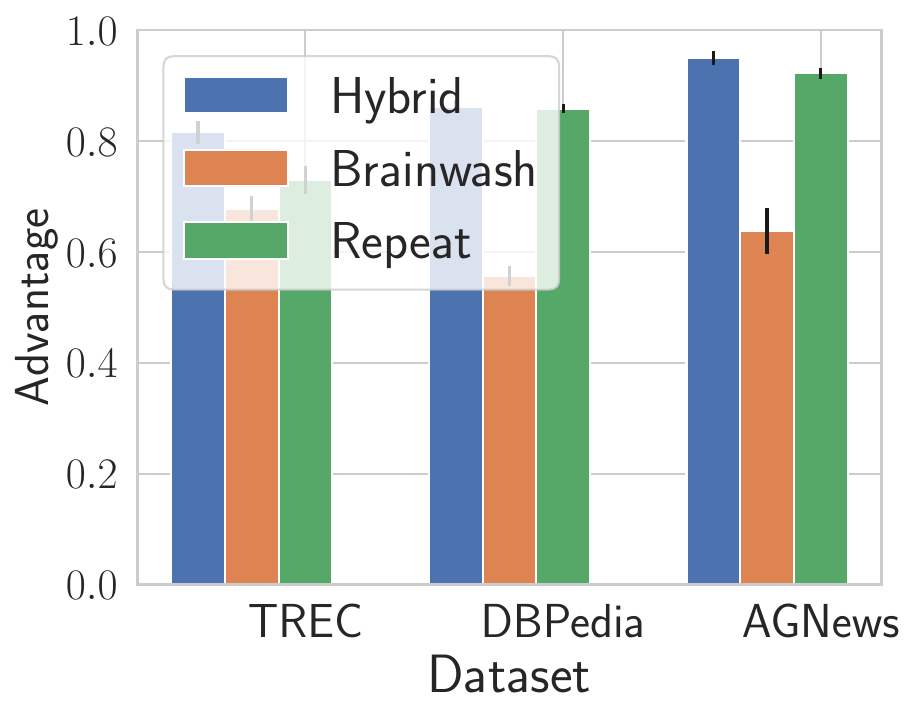}
\caption{GPT2-XL}
\label{fig:compare_hybrid_gpt2xl}
\end{subfigure}
\begin{subfigure}{0.49\columnwidth}
\includegraphics[width=\columnwidth]{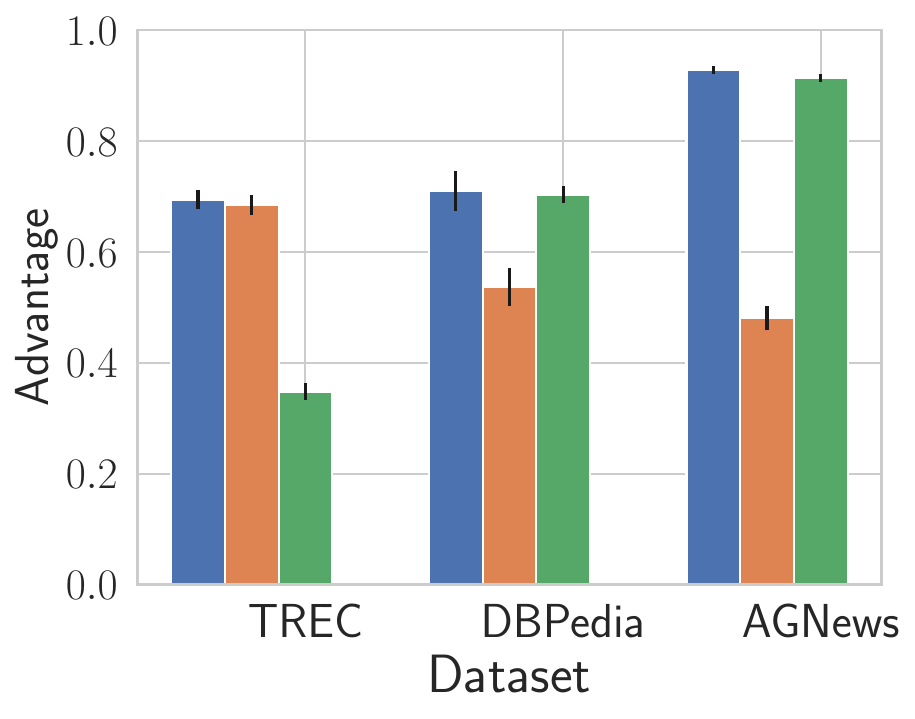}
\caption{GPT-3.5}
\label{fig:compare_hybrid_gpt35}
\end{subfigure}
\caption{Performance comparison of the hybrid attack against individual Brainwash and Repeat attacks across two language models. 
The hybrid attack effectively combines the strengths of both, often surpassing individual attack performances. 
In this figure, language models are prompted with one example. 
Full results on four language models can be found in~\refappendix{append:hybridresults}.}
\label{figure:compare_hybrid}
\end{figure}

Recall that the Repeat attack involves sending the first few words to the language model and assessing the semantic similarity between generated text and target samples, while the Brainwash attack iteratively induces the language model with incorrect answers and uses the number of iterations to determine membership status. 
In the hybrid attack, we concatenate both the similarity and iteration number, training a two-layer neural network as the attack model.

The attack model comprises fully connected layers and takes two inputs: the similarity returned by the Repeat attack and the average iteration number from the Brainwash attack. 
The output is the probability of membership.
To train the attack model, we assume the adversary can collect a small shadow dataset sampled from the same distribution as the demonstrations contained in $\prompt$. 
Our experiments indicate that training the attack model with the ensemble of Brainwash and Repeat attacks is not arduous: only 100 samples are sufficient.

Once the attack model is trained, the hybrid attack is deployed as follows: for a target sample, both Repeat and Brainwash attacks are applied to obtain corresponding metrics (similarity and iteration number). 
These values are then fed into the attack model to obtain the final prediction.

%-------------------------------------------------------------------------------
\subsection{Results}
%-------------------------------------------------------------------------------

We evaluate the effectiveness of our hybrid attack across four language models and present the results in \autoref{figure:compare_hybrid}. 
Notably, the hybrid attack capitalizes on the strengths of both Brainwash and Repeat attacks. 
In the majority of cases, the hybrid attack exhibits performance no less than the optimal performance achievable by Brainwash and Repeat attacks individually. 
Moreover, in certain scenarios, the hybrid attack even outperforms each individual attack. 
For instance, in \autoref{fig:compare_hybrid_gpt2xl}, where Brainwash and Repeat attacks achieve advantages of 67.8\% and 73.0\%, respectively, the hybrid attack achieves an advantage of 81.2\%, showcasing its ability to derive benefits from both attacks.

\begin{figure}[!t]
\centering
\includegraphics[width=0.66\columnwidth]{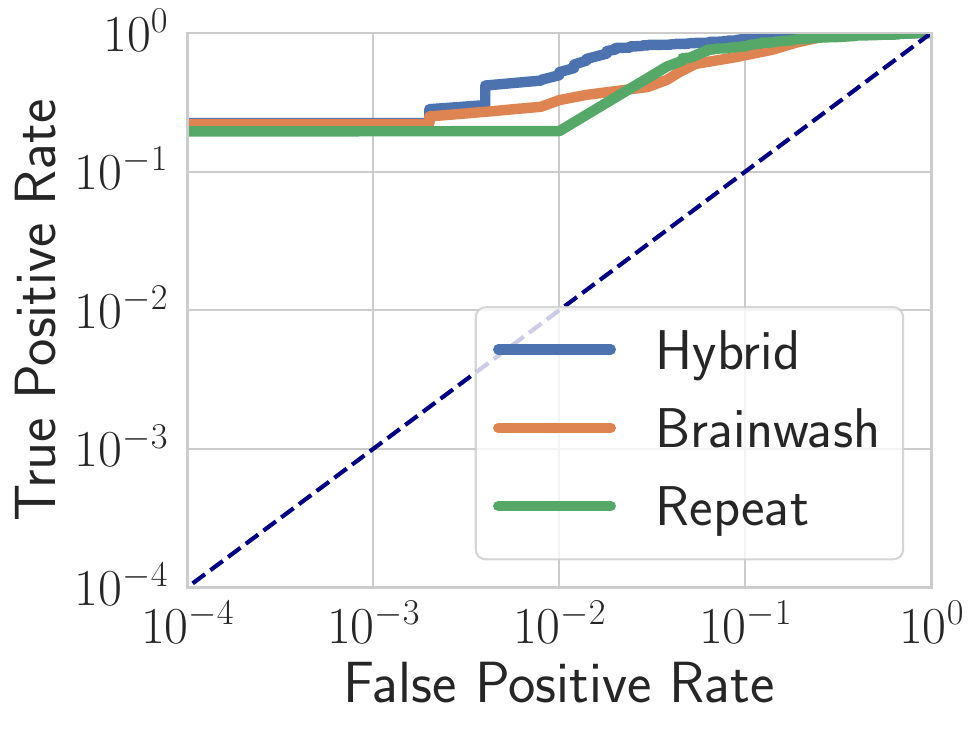}
\caption{Log-scale ROC curve illustrating the performance of the hybrid attack in leveraging the strengths of both Brainwash and Repeat attacks. 
The hybrid attack strategically combines their advantages to achieve superior performance across the false positive rate spectrum.}
\label{figure:hybrid_roc_one}
\end{figure}

To delve into how the hybrid attack leverages the advantages of both attacks, we present the log-scale ROC curve in \autoref{figure:hybrid_roc_one}. 
In the high false positive rate area, the hybrid attack capitalizes on the superior performance of the Repeat attack, while in the low false positive area, it aligns with the strategy of the Brainwash attack. 
This strategic combination results in high overall performance across the entire false positive rate area.

\begin{figure}[!t]
\centering
\begin{subfigure}{0.49\columnwidth}
\includegraphics[width=\columnwidth]{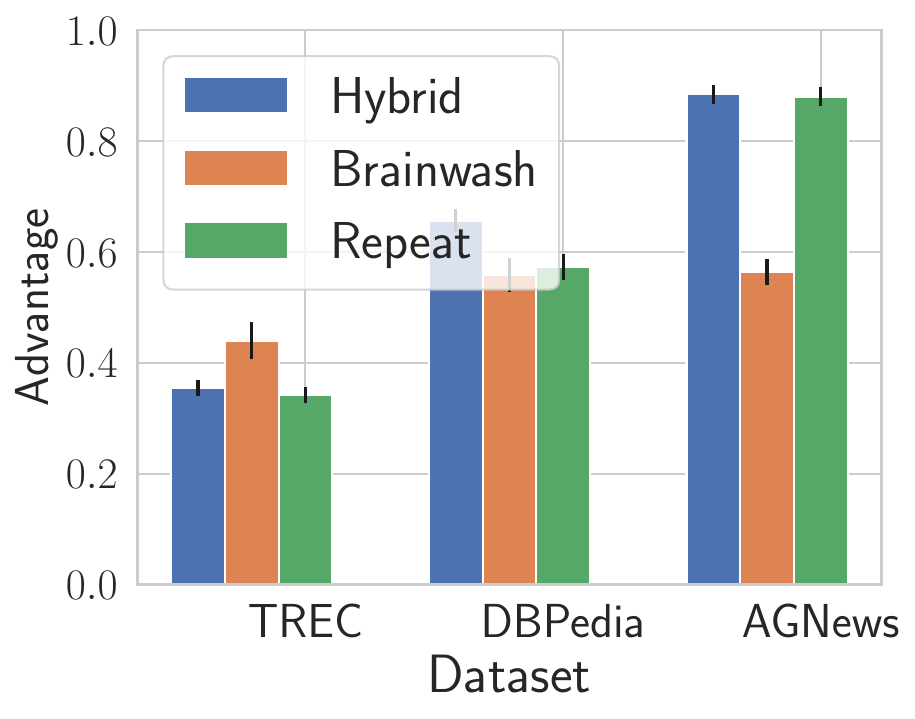}
\caption{Six Demonstrations (last)}
\label{fig:hybrid_six_end}
\end{subfigure}
\begin{subfigure}{0.49\columnwidth}
\includegraphics[width=\columnwidth]{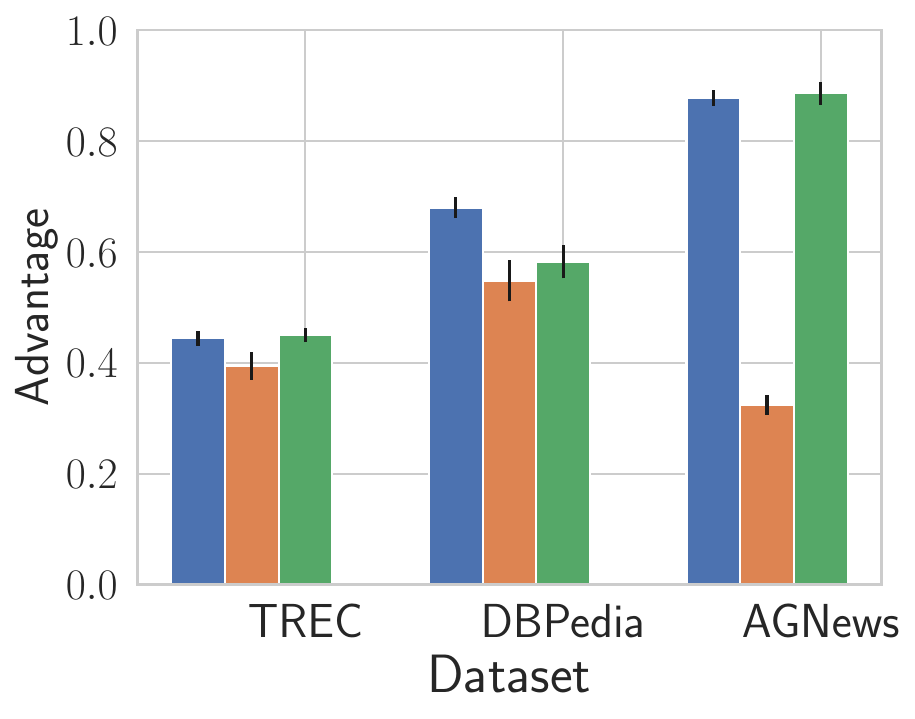}
\caption{Six Demonstrations (first)}
\label{fig:hybrid_six_begin}
\end{subfigure}
\caption{Performance evaluation of the hybrid attack on prompts with multiple demonstrations (\autoref{fig:hybrid_six_end}) and at various positions within a prompt (\autoref{fig:hybrid_six_begin}), demonstrating its consistent efficacy across different settings. 
In this figure, the language model used is Vicuna.}
\label{figure:hybrid_multi_position}
\end{figure}

Furthermore, we demonstrate that the hybrid attack maintains its advantage when targeting $\prompt$ consisting of multiple demonstrations (\autoref{fig:hybrid_six_end}) and attacking demonstrations at different positions within the $\prompt$ (\autoref{fig:hybrid_six_begin}). 
This evidence suggests that an adversary does not need to select a specific language model/dataset to attack; instead, the hybrid attack proves to be effective in different scenarios and can be used as a general attack against In-Context Learning.

%-------------------------------------------------------------------------------
\section{Potential Defenses}
%-------------------------------------------------------------------------------

Our proposed attack demonstrates effective performance in inferring the membership status of target samples, revealing significant privacy threats. 
However, as of our current knowledge, there is a lack of a comprehensive defense framework to safeguard In-Context Learning (ICL) from membership inference attacks.

In this section, we explore three potential defenses aiming to minimize the information leakage from the language model regarding its prompt.

\begin{figure*}[!t]
\centering
\begin{subfigure}{0.49\columnwidth}
\includegraphics[width=\columnwidth]{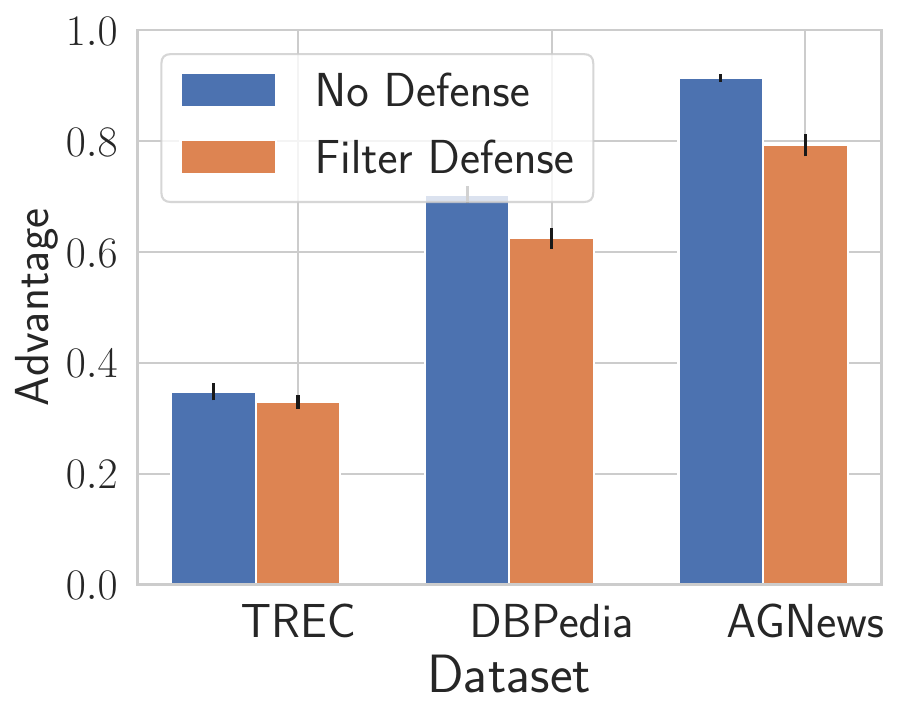}
\caption{One Demonstration}
\label{fig:filter_defense_one}
\end{subfigure}
\begin{subfigure}{0.49\columnwidth}
\includegraphics[width=\columnwidth]{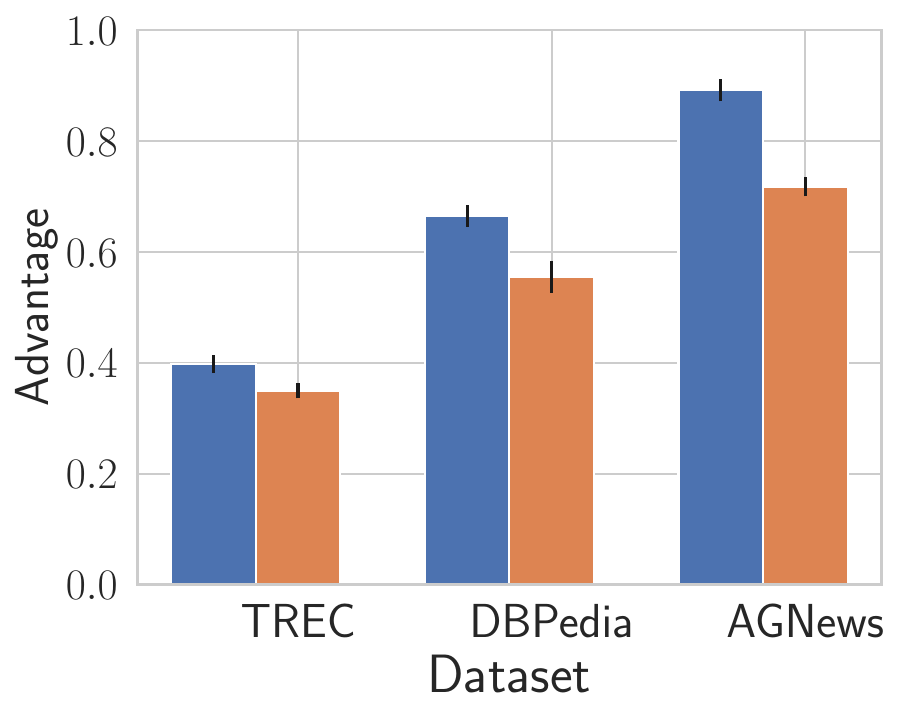}
\caption{Six Demonstrations}
\label{fig:filter_defense_six_end}
\end{subfigure}
\begin{subfigure}{0.49\columnwidth}
\includegraphics[width=\columnwidth]{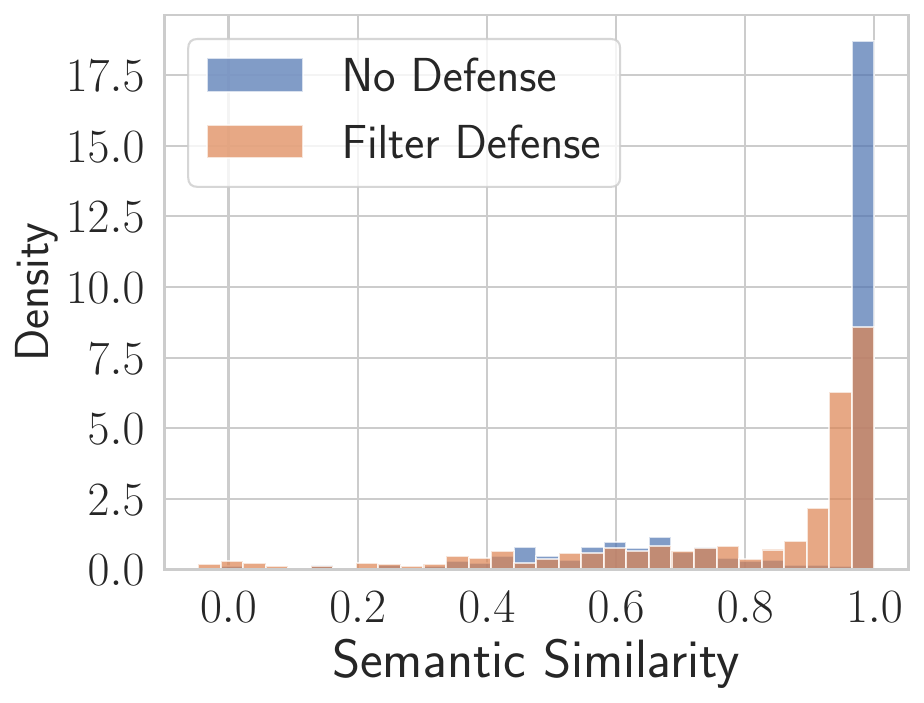}
\caption{Similarity Distribution}
\label{fig:filter_defense_dist}
\end{subfigure}
\begin{subfigure}{0.49\columnwidth}
\includegraphics[width=\columnwidth]{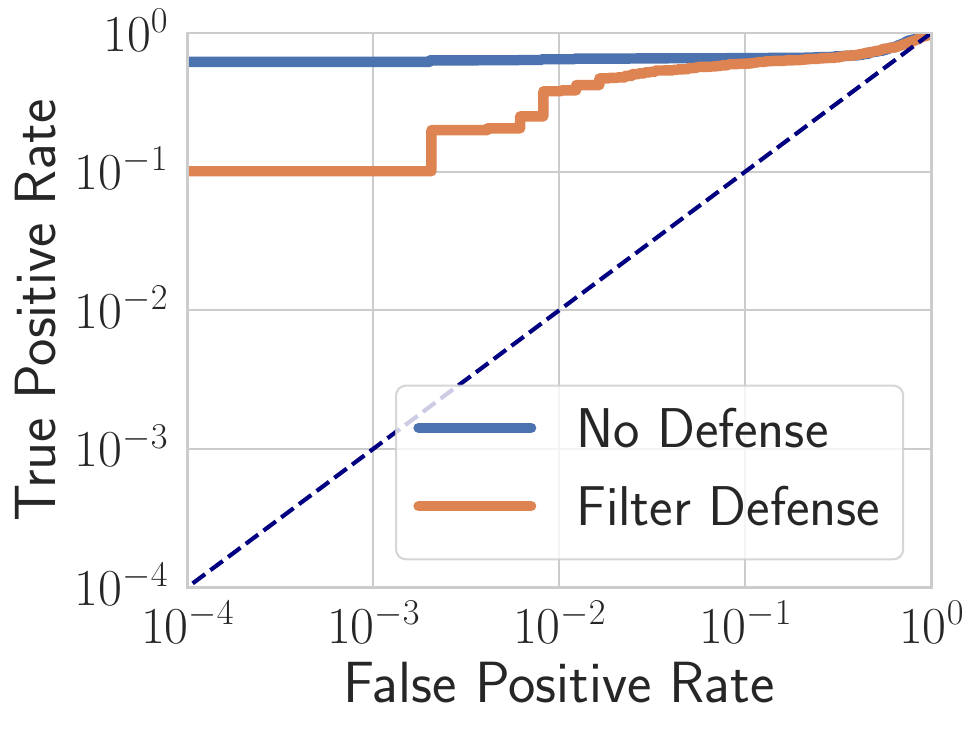}
\caption{Log-scale ROC Curve}
\label{fig:filter_defense_roc}
\end{subfigure}
\caption{Evaluation of a filter-based defense strategy against our attacks.
~\autoref{fig:filter_defense_one} and~\autoref{fig:filter_defense_six_end} demonstrate the defense's impact on Repeat attack performance. ~\autoref{fig:filter_defense_dist} depicts the change in semantic similarity distribution for member samples before and after the defense. ~\autoref{fig:filter_defense_roc} presents the log-scale ROC curve, highlighting the defense's effectiveness against worst-case performance scenarios.}
\label{figure:filter_defense_trec}
\end{figure*}

\begin{figure}[!t]
\centering
\begin{subfigure}{0.49\columnwidth}
\includegraphics[width=\columnwidth]{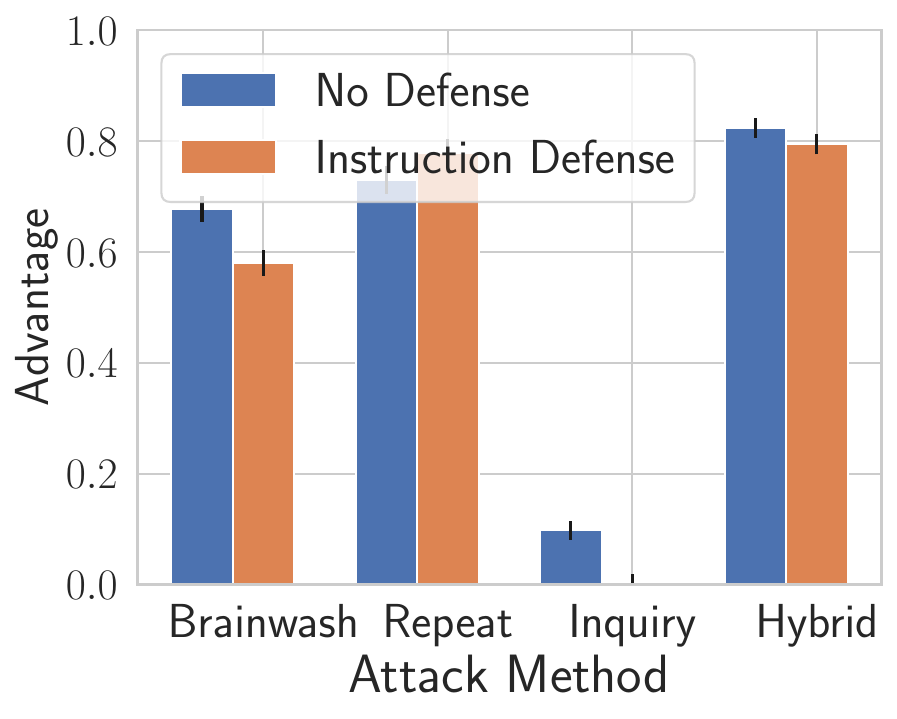}
\caption{GPT2-XL}
\label{fig:role_defense_trec_gpt2xl}
\end{subfigure}
\begin{subfigure}{0.49\columnwidth}
\includegraphics[width=\columnwidth]{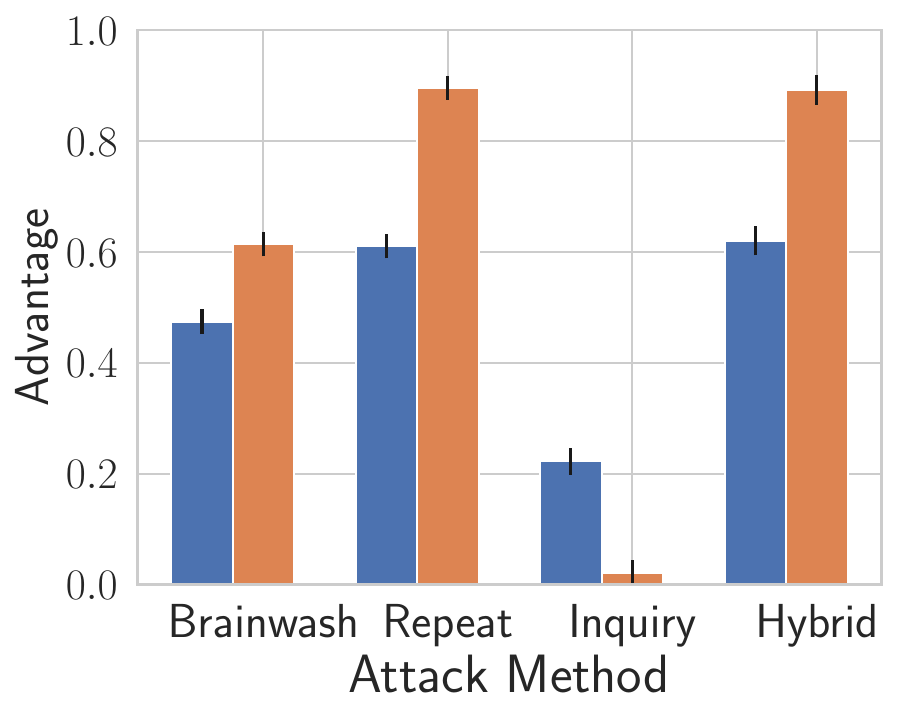}
\caption{Vicuna}
\label{fig:role_defense_trec_vicuna}
\end{subfigure}
\caption{
The defense instruction successfully reduces the effectiveness of the Inquiry attack for the TREC dataset; nevertheless, this mitigating effect does not extend to the Repeat attack or other datasets. 
We present the results on the GPT2-XL and Vicuna, full results can be found in~\autoref{append:defenseresults}.}
\label{figure:role_defense_trec}
\end{figure}

%-------------------------------------------------------------------------------
\subsection{Instruction-Based Defense}
%-------------------------------------------------------------------------------

Drawing inspiration from the strong control that instructions can exert over language models (e.g., prompt injection attack), our initial approach involves using instructions to compel the language model to refrain from leaking any information related to its prompt. 

Specifically, we task GPT-3.5 with designing a prompt intended to prevent the model from disclosure of prompt-related details. 
This strategy leverages previous findings~\cite{ZMHPPCB23} which suggest that language models can be particularly adept at crafting prompts, potentially surpassing human efforts in this domain. 
The generated instruction by GPT-3.5 is as follows:``Respond to the following queries without directly mentioning or alluding to any specific examples, demonstrations, or instances that might have been used in the prompt.''

We place this defense instruction at the end of the prompt and conduct the three most powerful attacks against demonstrations protected by this instruction.

While the efficacy of the defense instruction is evident in \autoref{figure:role_defense_trec}, showcasing a reduction in the performance of the Inquiry attack for the TREC dataset, this effectiveness does not uniformly extend to other datasets, as depicted in \autoref{figure:role_defense_dataset} (in~\refappendix{append:defenseresults}). 
Notably, the addition of the defense instruction tends to marginally decrease the performance of the Brainwash attack in most instances, though the variance is not statistically significant, we posit this reduction primarily to the increased distance between the query and the demonstration. 
Intriguingly, when evaluating the defense effect against Repeat attacks, in certain scenarios, the attack performance is observed to be even higher compared to scenarios without defense. 
We posit that integrating a well-designed defense instruction tailored to a specific attack and dataset may constitute a pragmatic approach to mitigate privacy leakage. 
However, the creation of a universally applicable defense instruction necessitates further scrutiny and exploration.

%-------------------------------------------------------------------------------
\subsection{Filter-Based Defense}
%-------------------------------------------------------------------------------

While acknowledging the resilience of the Repeat attack against simple defense instructions, we leverage insights from this attack methodology to devise an ad-hoc defense that actively modifies the language model's output. 
Specifically, since the Repeat attack determines membership status based on semantic similarity between the generated response and the target sample, we implement an output filter that modifies the response while preserving its utility. 
To achieve this, when the language model outputs content, we send that content to GPT-3.5 and request a sentence rewrite. 
This filter-based defense consistently reduces the performance of the Repeat attack across all datasets, as illustrated in both \autoref{fig:filter_defense_one} and \autoref{fig:filter_defense_six_end}.

\begin{figure*}[h]
\centering
\begin{subfigure}{0.58\columnwidth}
\includegraphics[width=\columnwidth]{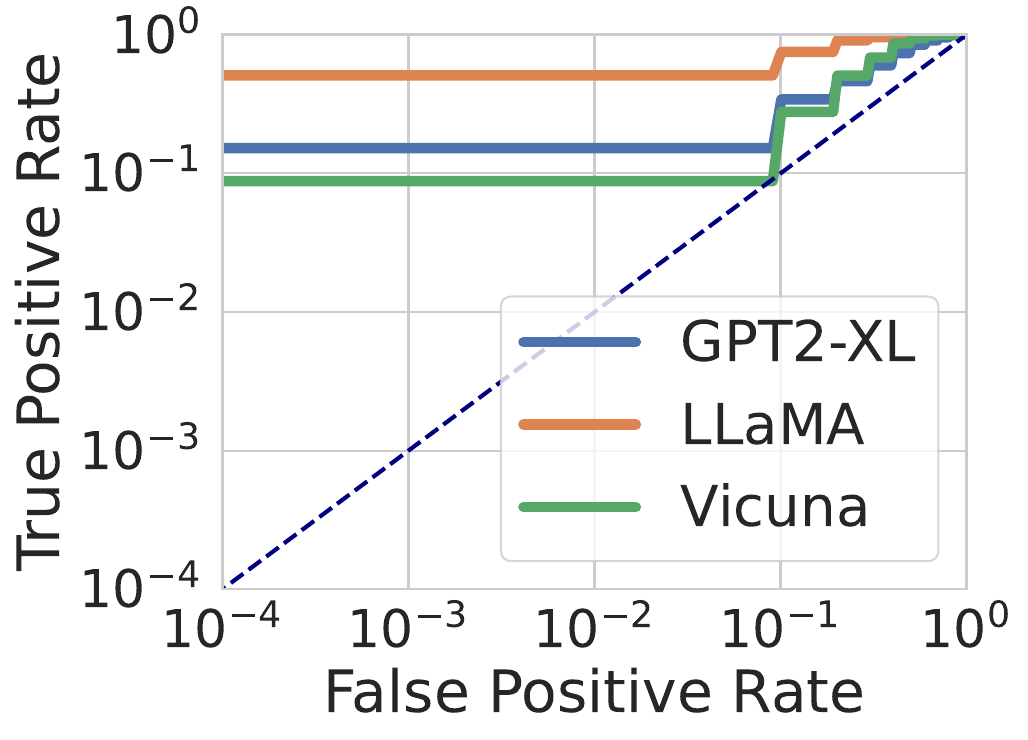}
\caption{TREC}
\label{fig:worst_case_trec}
\end{subfigure}
\begin{subfigure}{0.58\columnwidth}
\includegraphics[width=\columnwidth]{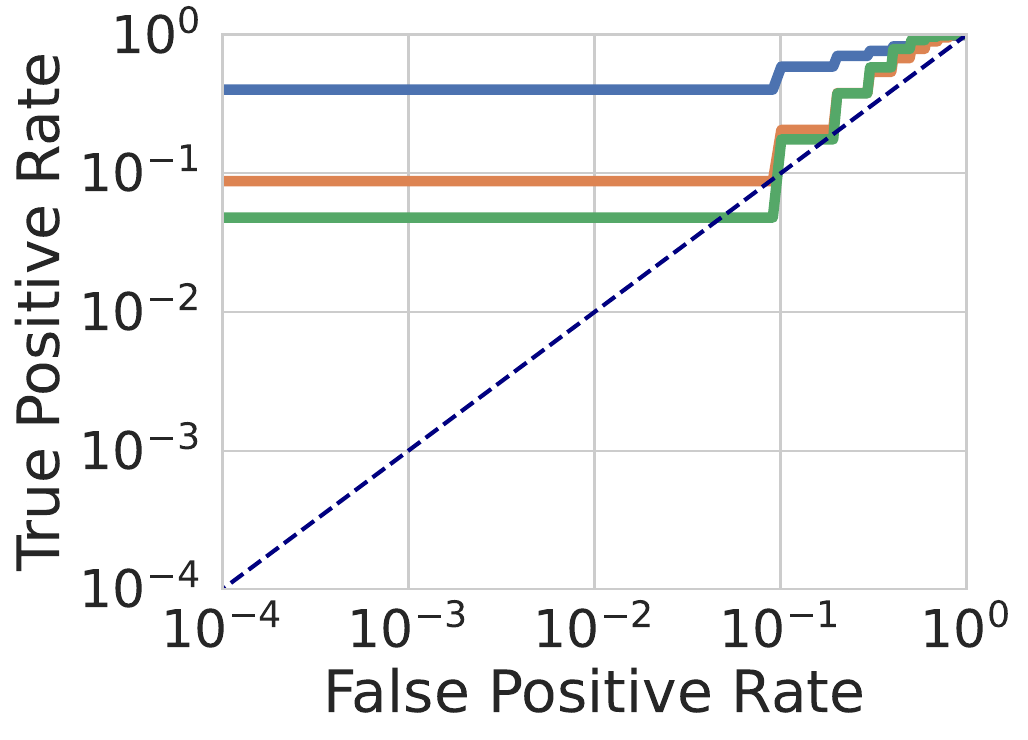}
\caption{DBPedia}
\label{fig:worst_case_dbpedia}
\end{subfigure}
\begin{subfigure}{0.58\columnwidth}
\includegraphics[width=\columnwidth]{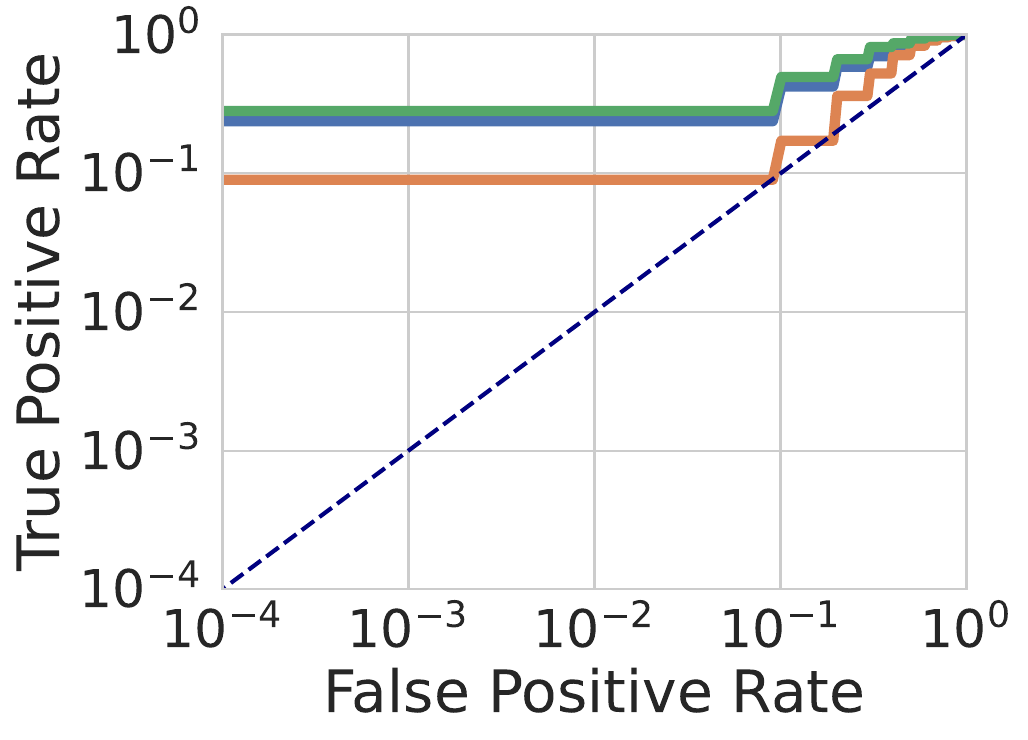}
\caption{AGNews}
\label{fig:worst_case_agnews}
\end{subfigure}
\caption{Worst-case evaluation with access to the posterior, these results provide an estimation of how traditional posterior-based attack could perform for the in-context learning paradigm.}
\label{figure:posterior_estimation}
\end{figure*}

\begin{figure}[t]
\centering
\includegraphics[width=0.68\columnwidth]{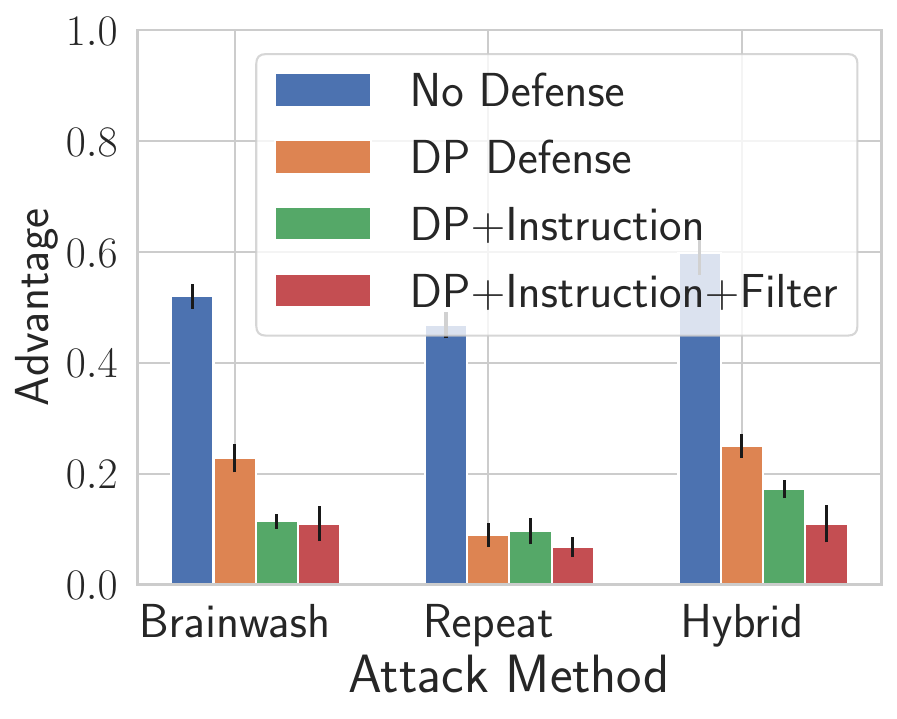}
\caption{Effectiveness of combining defenses from multiple dimensions. 
Results demonstrate that integrating defenses targeting different dimensions—such as data, instruction, and postprocessing—substantially enhances overall defense effectiveness and significantly reduces privacy leakage.}
\label{figure:dp_combine_defense}
\end{figure}

It is worth noting that our approach involves actively modifying the response, distinguishing it from common filter defenses~\cite{ZI23}. 
Blacklist-based filter defenses, which return an empty string if the output significantly overlaps with the prompt, may initially seem effective against prompt leakage but are susceptible to circumvention. 
For instance, an attacker could instruct the language model to output text encoded in a Caesar cipher or introduce additional spaces between characters to evade detection. 
On the other hand, whitelist-based filters, which only permit output from a predefined list, pose a greater challenge to bypass but may impact the utility, and we have considered and proposed the Brainwash attack tailored for this challenging scenario.

To understand how our filter-based defense diminishes the attack performance of Repeat, we analyze the semantic similarity distribution for member samples before and after the defense. 
As shown in \autoref{fig:filter_defense_dist}, before the defense, a considerable number of responses generated by the language model exhibit high similarity to the target sample (similarity close to 1). 
Our observations, along with previous work~\cite{ZI23}, indicate that the language model can sometimes reproduce the exact content from its prompt. 
However, after applying the output filter, the semantic similarity spreads more smoothly rather than concentrating near 1. 
The log-scale ROC curve (\autoref{fig:filter_defense_roc}) emphasizes that the performance degradation primarily occurs in the low false-positive rate area, indicating the defense's effectiveness against worst-case scenarios.

It's important to note that while rewriting the output is effective against the Repeat attack, it is not applicable to Brainwash and Inquiry attacks. 
These attacks only require the model to produce a predefined output or a binary answer, and manipulating the output can either distort the intended meaning (thus impacting utility) or render the defense ineffective. 
Consequently, the implementation of output filtering stands as a supplementary defense tailored to specific attack types, rather than offering a comprehensive, universal solution.

%-------------------------------------------------------------------------------
\subsection{DP-based Defense}
\label{sec:dp_defense}
%-------------------------------------------------------------------------------

Differential privacy (DP) has been established as a key defense mechanism against membership inference attacks. 
In this section, we assess the effectiveness of an existing DP-based defense strategy~\cite{TSIMMLGKS23}, which generates synthetic demonstrations from the private dataset with DP guarantees. 
Specifically, we set num-private-train to 1 to exclude unseen data. 
The resulting DP demonstrations are of low quality, such as ``Users Admin ApollosemblingIC negatives direct@GetMapping.'' 
Because of the high dissimilarity between the generated DP and original demonstrations, the effectiveness of the Repeat attack is reduced to almost random guessing.

However, DP is less effective against the Brainwash attack. 
Despite the generated samples being dissimilar to the original samples, the Brainwash attack still achieves a 0.228 advantage, as depicted in the yellow bar in~\autoref{figure:dp_combine_defense}.

As DP-based, instruction-based, and filter-based defenses target orthogonal components of the language model—namely, the data, the instruction, and the output—a logical approach is to combine these three defenses for enhanced protection.
In~\autoref{figure:dp_combine_defense}, we illustrate the synergy of combining different defense strategies.
On the basis of the DP-based defense, the instruction-based defense further reduces the performance of the Brainwash attack while having a negligible influence on the Repeat attack performance. 
Adding the filter-based defense on top of these two defenses further reduces the effectiveness of the Repeat attack. 
Upon combining all three defenses, the overall performance of the hybrid attack is reduced from 0.59 to 0.11. 
These results suggest that effective defense strategies should not focus on a single component but rather leverage a combination of defenses targeting orthogonal components.

%-------------------------------------------------------------------------------
\section{Discussion and Limitations}
\label{sec:discussion}
%-------------------------------------------------------------------------------

In this section, we discuss how our findings can advance our understanding of vulnerabilities in ICL, as well as the current limitations.

\mypara{For Attack}
We first evaluated the worst-case performance using a posterior-based method across three datasets for three open-source models: GPT2-XL, LLaMA, and Vicuna.
The results, presented in~\autoref{figure:posterior_estimation}, illustrate that utilizing posterior probabilities to determine membership status yields robust performance. 
This supports our hypothesis that samples within the prompt exhibit a significantly lower loss, indicating higher model confidence in demonstrations.

However, despite the effectiveness of the posterior-based attack, the inability to observe the loss directly from text output highlights the difficulty of conducting membership inference attacks using only text data. 
The main challenge lies in converting these unobservable aspects into observable ones.

\begin{figure}[!t]
\centering
\begin{subfigure}{0.49\columnwidth}
\includegraphics[width=\columnwidth]{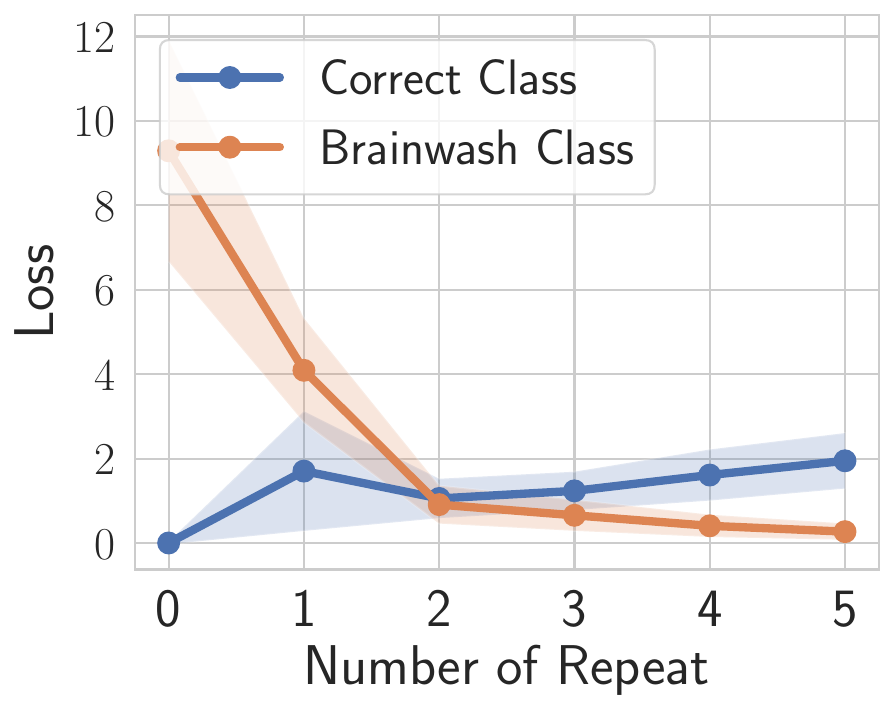}
\caption{Member}
\label{fig:brainwash_loss_member}
\end{subfigure}
\begin{subfigure}{0.49\columnwidth}
\includegraphics[width=\columnwidth]{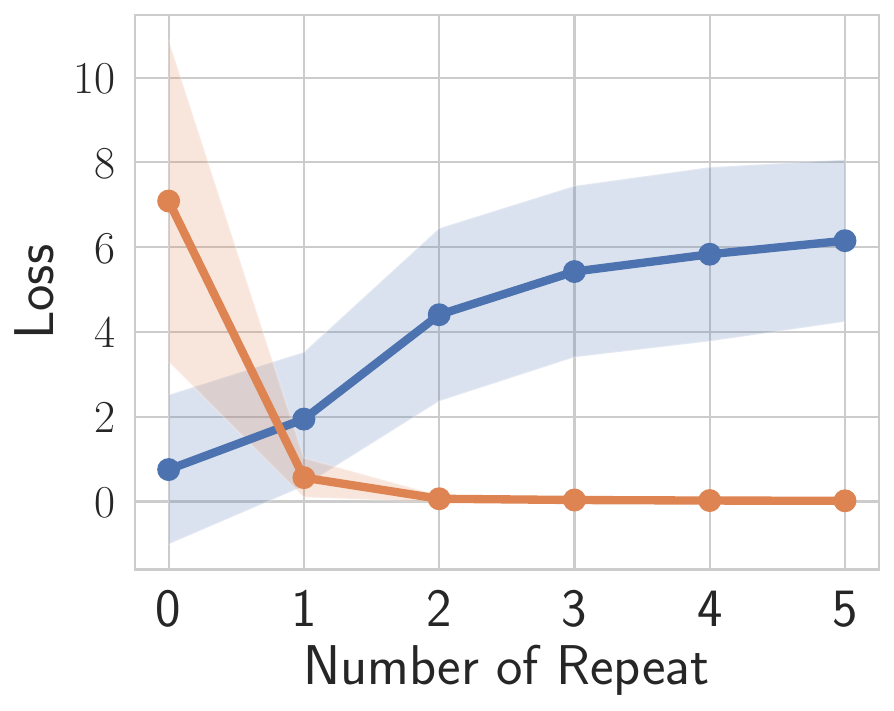}
\caption{Non-Member}
\label{fig:brainwash_loss_nonmember}
\end{subfigure}
\caption{Visualization of loss dynamics in the Brainwash attack: The figures demonstrate that the loss for the correct class increases when the language model is brainwashed by wrong labels, translating unobservable loss changes into observable signals, such as the number of repetitions.}
\label{figure:brainwash_loss}
\end{figure}

To tackle this, we conducted an in-depth analysis to understand how the Brainwash attack converts unobservable signals, such as loss, into observable ones.
Specifically, we visualized the \textit{loss dynamics} for both the correct class and the maliciously introduced ``brainwash'' class, as illustrated in \autoref{figure:brainwash_loss}.

Both two figures indicate that when we brainwash the language model using incorrect labels, the loss on the correct class gradually increases, while the loss on the incorrect brainwash class decreases. When the loss of the correct class surpasses that of the brainwash class, the language model predicts the wrong label. 
This phenomenon explains how ICL interacts with brainwash samples.

Furthermore, we can see that for member samples and non-member samples, the speed loss increase is different. 
Specifically, as shown in~\autoref{fig:brainwash_loss_nonmember}, for non-member samples that are not included in the prompt,  the loss on the correct class increases rapidly when facing brainwash samples.
This indicates that the language model's confidence in target samples can easily be influenced, and the language model quickly accepts the brainwash class, evidenced by the sharp decrease of the loss on the brainwash class.
However, for member samples, the loss on the correct class increases more slowly, and the brainwash class requires more repetitions to reduce the loss, resulting in a delayed intersection point.

This analysis highlights how our brainwash attack can transform unobservable signals, such as loss, into observable signals, such as the number of repetitions. 
This novel conversion makes the attack feasible in realistic and challenging scenarios where the target model outputs only text.
Besides, this approach is analogous to label-only membership inference attacks in the vision domain, where adversarial perturbations indicate model confidence in target samples. 
Our method can be regarded as an adversarial perturbation technique for LLMs with text-only output, even with restricted output.

We acknowledge that currently, it’s challenging to provide a theoretical understanding of the vulnerability, as the community is still exploring how ICL interacts with the model. 
For example, NeedleInAHayStack~\cite{NeedleInAHaystack} demonstrates that when facing a long context, the models are more likely to memorize information encoded at the beginning/end, but Anthropic~\cite{claude-2-1-prompting} later pointed out that these findings may be influenced by the prompt.

Our attack provides empirical evidence on how models memorize demonstrations in an adversarial setting and shares some common vulnerabilities, like the first and last demonstrations are more more vulnerable to privacy leakage. 
Moreover, our analysis suggests further attacks could follow the same intuition by proposing different methods that better approximate the confidence of the LLMs to the target samples, thereby enhancing attack performance.

\mypara{For Defense}
Our work provides insights into defense strategies, indicating that effective defenses should combine multiple components, such as data, instruction, and output, rather than focusing on a single aspect. 
As shown in~\autoref{figure:dp_combine_defense}, different defense mechanisms each have their own advantages, and combining defenses from orthogonal dimensions can result in better synergy.

Additionally, our study indicates that relying solely on developer interventions, such as Reinforcement Learning from Human Feedback (RLHF), may introduce side effects. 
This is evidenced by the varying attack performance across different model versions, as shown in~\autoref{figure:overtime}. 
A version that is effective at defending against one type of attack may inadvertently increase the model's vulnerability to another type of attack.

While it is challenging to draw a formalized conclusion at this stage, we believe our findings offer valuable insights that could benefit the community in better understanding these vulnerabilities.

%-------------------------------------------------------------------------------
\section{Ethical and Privacy Considerations}
\label{sec:ethical}
%-------------------------------------------------------------------------------

Membership inference attacks against ICL in LLMs pose significant privacy risks, as they can reveal whether specific data points were used in the model's demonstrations. 
This threat can lead to the exposure of sensitive personal information, undermining user trust and potentially causing harm. 
Ethically, it is crucial to ensure transparency, fairness, and accountability in the use of LLMs. 

On the other hand, these attacks can serve as auditing tools to verify whether unauthorized data has been used to construct prompts.
To mitigate these risks, we suggest a series of strategies, including differential privacy, instruction-based, and filter-based defenses. 
We strongly recommend that developers integrate these approaches when releasing query APIs. 
Employing these mitigation strategies can protect user data and maintain ethical standards in AI development and deployment.

%-------------------------------------------------------------------------------
\section{Conclusion}
\label{sec:conclusion}
%-------------------------------------------------------------------------------

In this paper, we propose the first text-only membership inference attack against ICL. 
Our attack exhibits effectiveness across various scenarios, including instances where the language model is constrained to generating responses from a predefined list. 
We conduct extensive experiments across diverse datasets and language models and empirically demonstrate the effectiveness of our attacks.
We delve into an exploration of factors influencing attack efficacy, revealing that the vulnerability of demonstrations results from the intricate interplay between prompt size and the demonstration position.
A thorough investigation into the information leakage in language models over time uncovers persistent vulnerabilities even with updated versions, heightening our concerns. 
To mitigate membership leakage, we explore three potential defenses, finding that their combination significantly reduces privacy leakage.

Our study not only enhances our understanding of the intricacies of ICL vulnerabilities but also contributes practical considerations for prompt design and defense mechanisms. 
Despite the successful implementation of defenses in specific scenarios, the quest for a comprehensive and generalized defense strategy persists. 
As the field continues to advance, our findings provide a foundation for researchers and practitioners alike, guiding efforts toward a more secure and privacy-conscious integration of ICL into the transformative landscape of LLMs.

\section*{Acknowledgements}
We thank all anonymous reviewers for their constructive comments.
This work is partially funded by the European Health and Digital Executive Agency (HADEA) within the project ``Understanding the individual host response against Hepatitis D Virus to develop a personalized approach for the management of hepatitis D'' (DSolve, grant agreement number 101057917) and the BMBF with the project ``Repräsentative, synthetische Gesundheitsdaten mit starken Privatsphärengarantien'' (PriSyn, 16KISAO29K).

\bibliographystyle{plain}
\bibliography{normal_generated_py3,blog}

\clearpage
\appendix

%-------------------------------------------------------------------------------
\section{Influence of Memorization on Attack Performance}
\label{append:memorization_influence}
%-------------------------------------------------------------------------------

\begin{figure}[h]
\centering
\includegraphics[width=0.66\columnwidth]{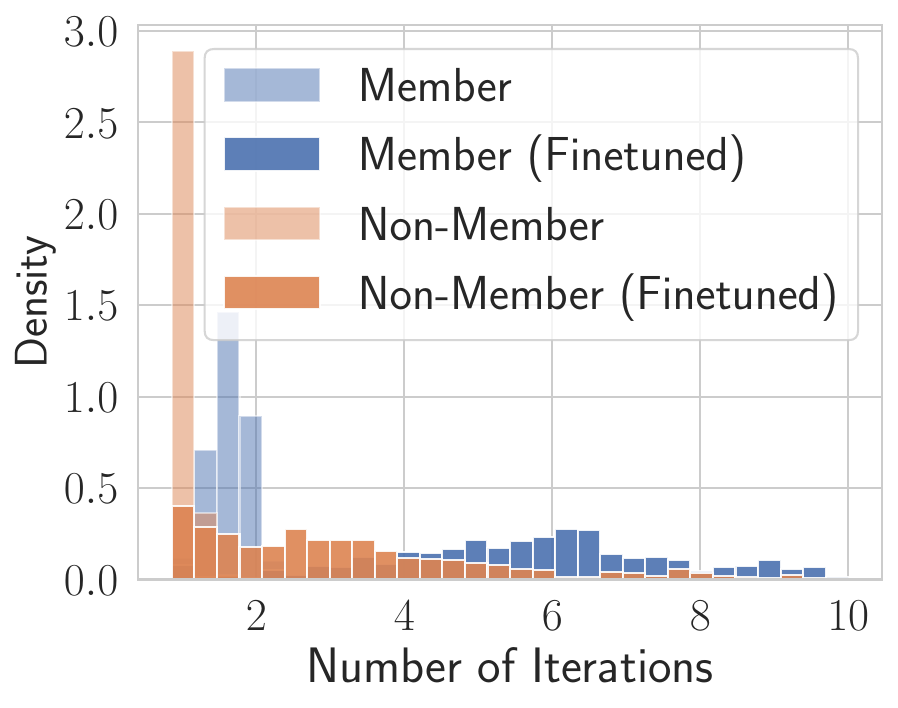}
\caption{Distribution of the number of iterations before and after fine-tuning. 
The results indicate that post-fine-tuning, the model exhibits increased confidence in its samples, irrespective of their membership status. 
Both member and non-member samples require a greater number of iterations to alter their predictions, leading to a degradation in attack performance. 
The experiments are conducted on the DBPedia dataset using the LLaMA model.}
\label{figure:memorization_influence}
\end{figure}

To investigate the effect of memorization, we fine-tuned the LLaMA model using the LoRA technique with 2,000 samples from DBPedia, resulting in the model heavily memorizing DBPedia content. 
The attack performance showed a significant decrease, with accuracy dropping from 0.934 to 0.782. 
Specifically, after being fine-tuned on the DBPedia dataset, the model exhibited increased confidence in its responses to test samples, regardless of their presence in the prompt. 
As illustrated in~\autoref{figure:memorization_influence}, after fine-tuning, more number of brainwash is required to change the model's prediction, indicating strong memorization.

However, it is important to note that this represents an extreme case. 
Our experiments are conducted on commonly used benchmark datasets typically used for pre-training. 
If evaluated on entirely unseen datasets, we would expect even better performance, as the model would be less confident on unseen data.

%-------------------------------------------------------------------------------
\section{Attack Performance Over Time (DBPedia)}
\label{append:overtime_dbpedia}
%-------------------------------------------------------------------------------

\begin{figure}[h]
\centering
\includegraphics[width=0.66\columnwidth]{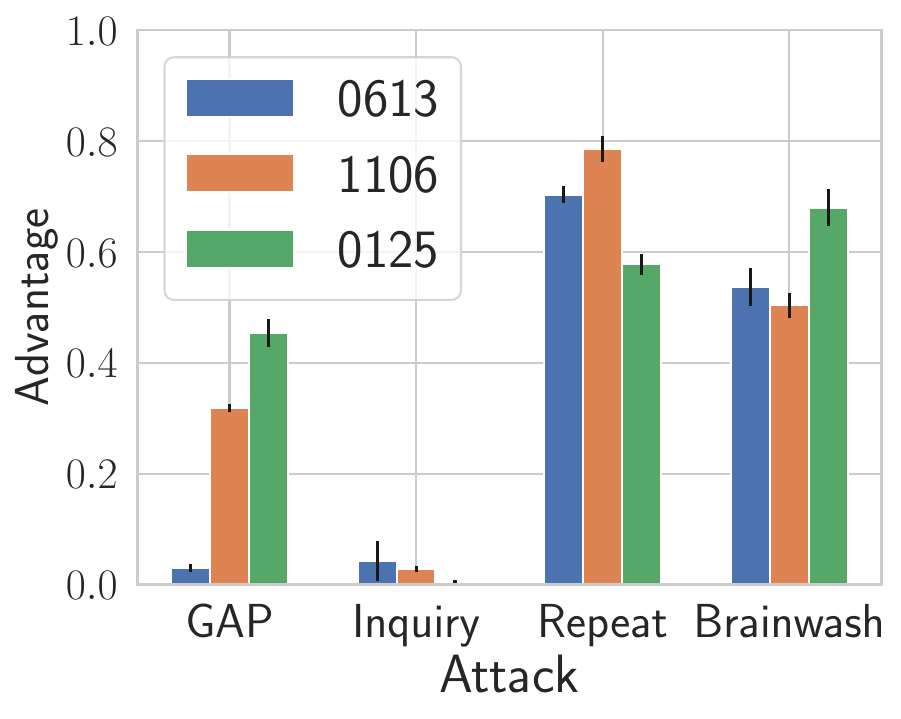}
\caption{We evaluate the evolution of attack performance on the DBPedia dataset using three versions of the GPT-3.5 API (gpt-3.5-turbo-0613, gpt-3.5-turbo-1106, and gpt-3.5-turbo-0124). 
The results from the DBPedia dataset align with our findings on the TREC dataset, indicating that the robustness of commercial models like GPT-3.5 does not consistently improve over time.}
\label{figure:overtime_dbpedia}
\end{figure}

We investigated performance changes using the DBPedia dataset. 
Due to the deprecation of some versions, we conducted experiments on gpt-3.5-turbo-0613, gpt-3.5-turbo-1106, and gpt-3.5-turbo-0125. 
Results presented in \autoref{figure:overtime_dbpedia} indicate that no single version consistently outperforms the others in terms of robustness.

%-------------------------------------------------------------------------------
\section{Long Context Performance}
\label{append:longcontext}
%-------------------------------------------------------------------------------

\begin{figure}[!t]
\centering
\includegraphics[width=0.66\columnwidth]{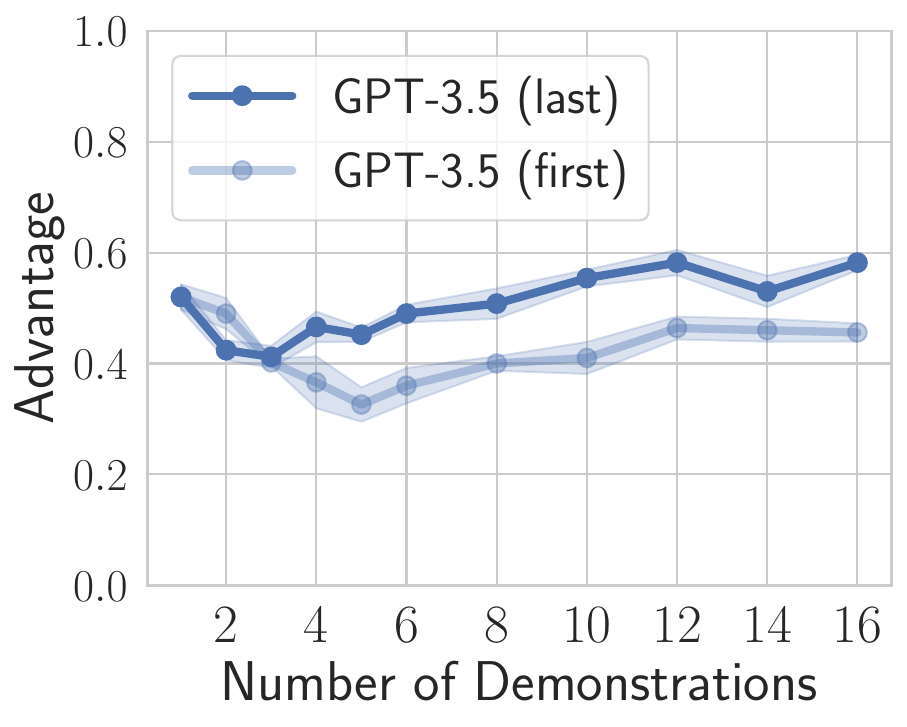}
\caption{Comparative analysis of Brainwash performance targeting the first and last demonstration, with the number of demonstrations ranging from 1 to 16. 
The experiments utilize the gpt-3.5-turbo-0125 model and are performed on the TREC dataset.}
\label{figure:more_demo}
\end{figure}

As an increasing number of language models support long-context capabilities, it is important to understand how vulnerabilities evolve with extended demonstrations. 
In this section, we experiment with the latest version of GPT-3.5 (gpt-3.5-turbo-0125), using a range of demonstrations from 1 to 16, and present the results in \autoref{figure:more_demo}.

The results indicate that our attacks maintain high performance even as the number of demonstrations increases. 
Specifically, when using 16 TREC samples, the Brainwash attack achieves a 0.582 advantage when inferring the last demonstration, and this advantage remains at 0.456 for the first demonstration. 
We also compared these results on the latest version with our previous findings on gpt-3.5-turbo-0613. 
It appears that the latest model is more vulnerable to our attack when more demonstrations are included in the prompt. 
This may suggest that the latest version has enhanced capabilities for handling and remembering long contexts, reflecting the ongoing trend of large language models improving their long-context abilities.

\section{Hybrid Attack Results}
\label{append:hybridresults}

\begin{figure*}[!t]
\centering
\begin{subfigure}{0.49\columnwidth}
\includegraphics[width=\columnwidth]{revision_figs/hybrid_compare_arch_gpt2-xl_demo_1_position_end.pdf}
\caption{GPT2-XL}
\label{fig:compare_hybrid_gpt2xl_append}
\end{subfigure}
\begin{subfigure}{0.49\columnwidth}
\includegraphics[width=\columnwidth]{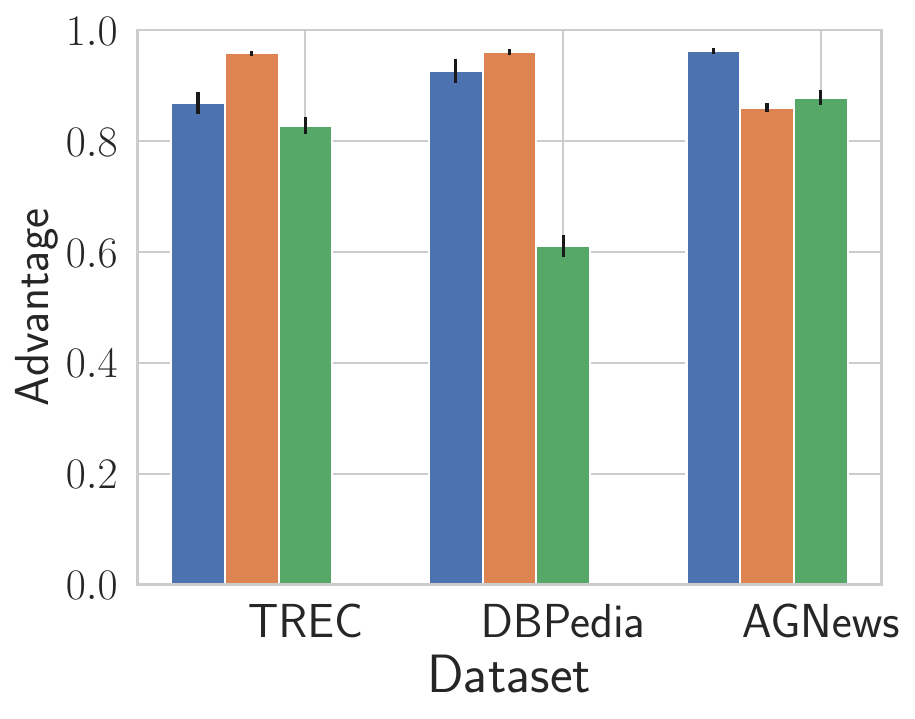}
\caption{LLaMA}
\label{fig:compare_hybrid_llama_append}
\end{subfigure}
\begin{subfigure}{0.49\columnwidth}
\includegraphics[width=\columnwidth]{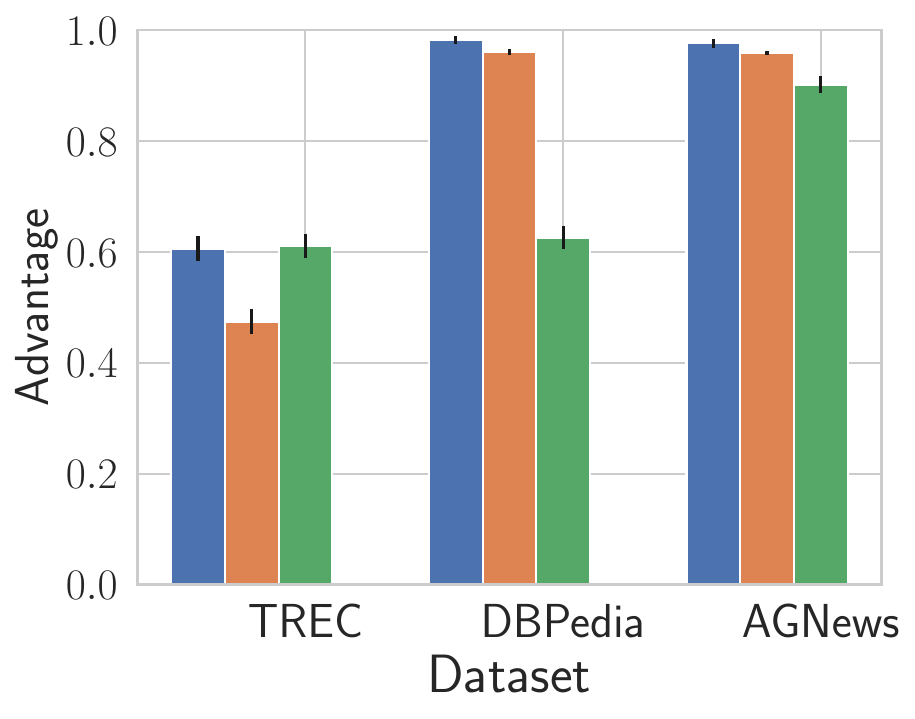}
\caption{Vicuna}
\label{fig:compare_hybrid_vicuna_append}
\end{subfigure}
\begin{subfigure}{0.49\columnwidth}
\includegraphics[width=\columnwidth]{revision_figs/hybrid_compare_arch_gpt3.5_demo_1_position_end.pdf}
\caption{GPT-3.5}
\label{fig:compare_hybrid_gpt35_append}
\end{subfigure}
\caption{Performance comparison of the hybrid attack against individual Brainwash and Repeat attacks across four language models. 
The hybrid attack effectively combines the strengths of both, often surpassing individual attack performances. 
In this figure, language models are prompted with one example.}
\label{figure:compare_hybrid_append}
\end{figure*}

We show more results of our hybrid attack in~\autoref{figure:compare_hybrid_append}. 
As shown in the figures, the hybrid attack effectively combines the strengths of both, often surpassing individual attack performances. 

%-------------------------------------------------------------------------------
\section{Defense Results}
\label{append:defenseresults}
%-------------------------------------------------------------------------------

\begin{figure*}[!t]
\centering
\begin{subfigure}{0.49\columnwidth}
\includegraphics[width=\columnwidth]{revision_figs/role_base_defense_dataset_trec_model_gpt2-xl_demo_1_position_end.pdf}
\caption{GPT2-XL}
\label{fig:role_defense_trec_gpt2xl_append}
\end{subfigure}
\begin{subfigure}{0.49\columnwidth}
\includegraphics[width=\columnwidth]{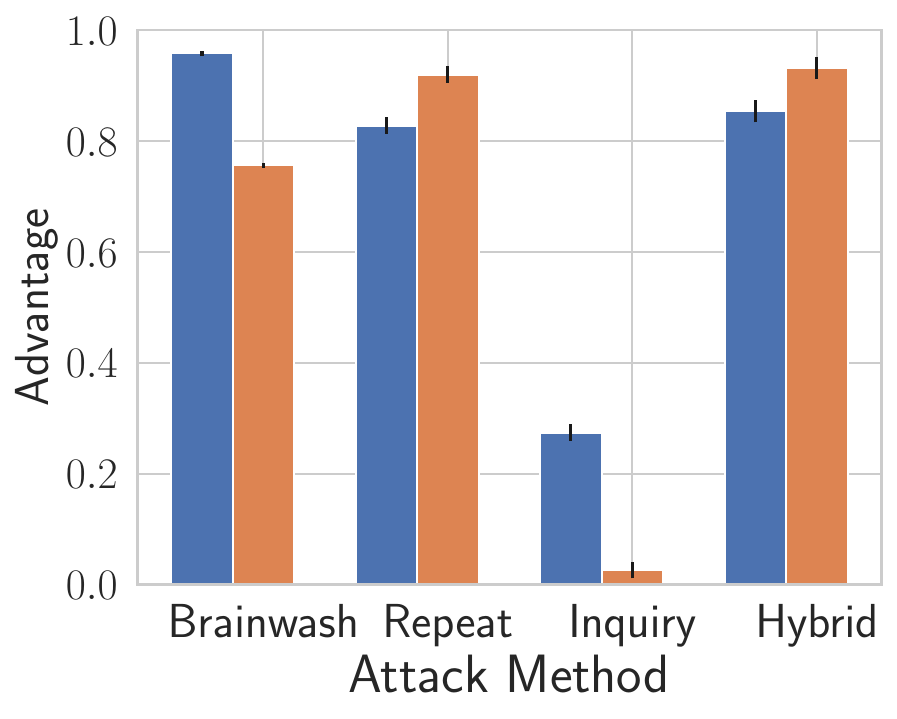}
\caption{LLaMA}
\label{fig:role_defense_trec_llama_append}
\end{subfigure}
\begin{subfigure}{0.49\columnwidth}
\includegraphics[width=\columnwidth]{revision_figs/role_base_defense_dataset_trec_model_vicuna_demo_1_position_end.pdf}
\caption{Vicuna}
\label{fig:role_defense_trec_vicuna_append}
\end{subfigure}
\begin{subfigure}{0.49\columnwidth}
\includegraphics[width=\columnwidth]{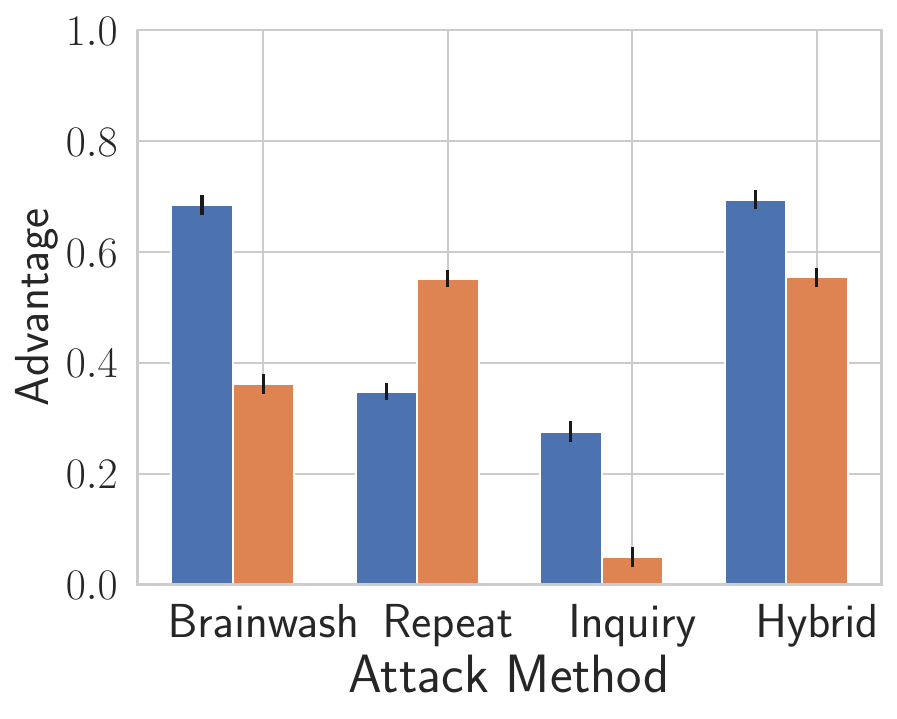}
\caption{GPT-3.5}
\label{fig:role_defense_trec_gpt35_append}
\end{subfigure}
\caption{
The defense instruction successfully reduces the effectiveness of the Inquiry attack for the TREC dataset; nevertheless, this mitigating effect does not extend to the Repeat attack or other datasets.}
\label{figure:role_defense_trec_append}
\end{figure*}

\begin{figure}[!t]
\centering
\begin{subfigure}{0.49\columnwidth}
\includegraphics[width=\columnwidth]{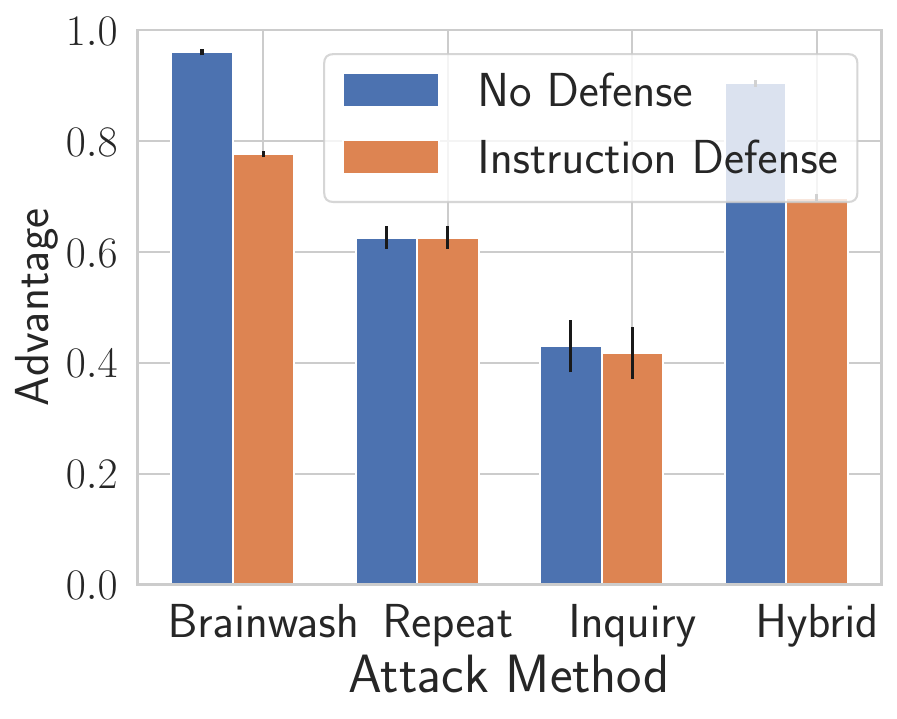}
\caption{DBPedia}
\label{fig:role_defense_dbpedia}
\end{subfigure}
\begin{subfigure}{0.49\columnwidth}
\includegraphics[width=\columnwidth]{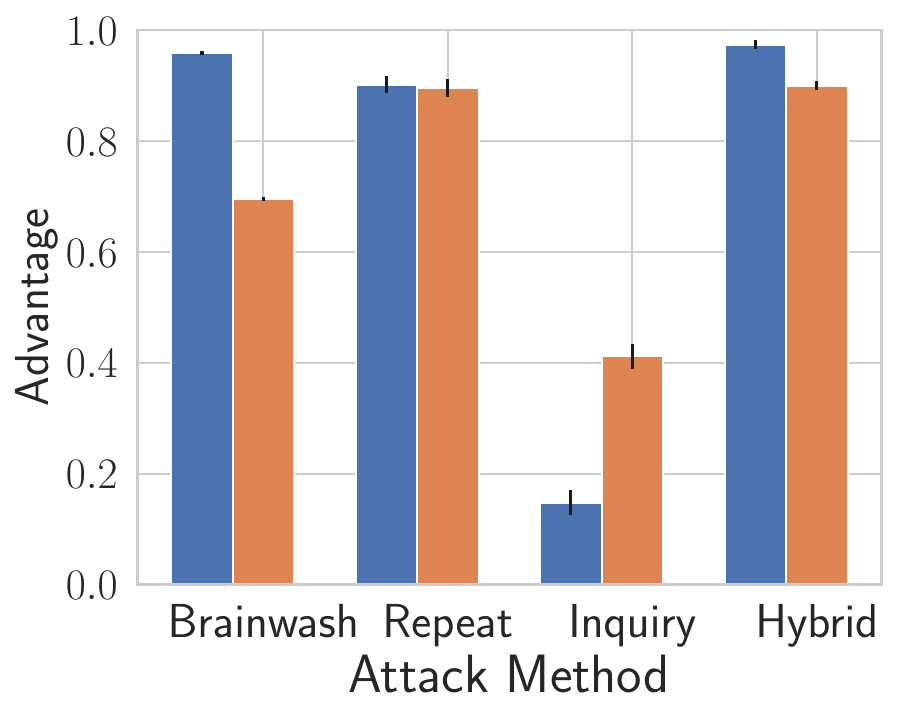}
\caption{AGNews}
\label{fig:role_defense_agnews}
\end{subfigure}
\caption{Evaluation of the defense instruction's impact on the performance of different attacks across diverse datasets, highlighting varying levels of efficacy and nuances in defense outcomes.}
\label{figure:role_defense_dataset}
\end{figure}

The defense instruction successfully reduces the effectiveness of the Inquiry attack for the TREC dataset, as shown in~\autoref{figure:role_defense_trec_append}; nevertheless, this mitigating effect does not extend to the Repeat attack or other datasets, as shown in~\autoref{figure:role_defense_dataset}.

%-------------------------------------------------------------------------------
\section{Related Work}
\label{sec:related}
%-------------------------------------------------------------------------------

%-------------------------------------------------------------------------------
\subsection{Privacy Attacks Against In-Context Learning}
%-------------------------------------------------------------------------------

One prominent privacy concern regarding In-Context Learning (ICL) involves the prompt extraction attack, designed to recover the prompt through API access to the language model.
Zhang and Ippolito~\cite{ZI23} first formulate a text-based extract attack. 
Contrary to conventional reconstruction attacks on vision models that optimize the input to reduce testing loss, their work reconstructs the prompt by sending instructions to the model without using backpropagation. 
This approach extends the applicability of the attack to black-box models, encompassing GPT-3.5 and GPT-4. 
However, the attack's performance is closely tied to the quality of instructions, and defensive measures, such as output filtering, may impede the attack.

Another category of attacks aims to identify whether specific samples were used to construct the prompt, known as membership inference attacks. 
Current work predominantly employs loss-based attacks, assuming the adversary can access the probability associated with the generated content. 
This allows the calculation of loss for the target sample to determine membership status. 
Duan et al.~\cite{DDYPB23} compare membership vulnerabilities for fine-tuning and ICL, demonstrating that ICL is more susceptible to membership inference attacks. 
Wen et al.~\cite{WWBZS23} further extend this privacy vulnerability comparison to other adaptation methods, including Low-Rank Adaptation (LoRA) and Soft Prompt Tuning (SPT), concluding that ICL exhibits the highest membership vulnerability among them. 
However, existing membership inference attacks lack the ability to operate under a text-only setting, leaving the vulnerability in a text-only scenario unexplored.

%-------------------------------------------------------------------------------
\subsection{Membership Inference Attacks}
%-------------------------------------------------------------------------------

Membership inference attacks~\cite{SSSS17,SZHBFB19,HWWBSZ21,LZ21,LZBZ22,HLXCZ22, WYLBZ22} aim to identify whether a sample has contributed to enhancing the utility of a model. 
These attacks not only serve as audit tools for data provenance but can also assist in reconstructing datasets. 
For instance, Carlini et al.~\cite{CTWJHLRBSEOR21} leverage language models to generate candidate samples, subsequently employing membership inference attacks to filter out non-member samples, achieving a highly effective reconstruction attack.

Carlini et al.\cite{CLEKS19} utilized perplexity as a metric, highlighting that member samples typically exhibit low perplexity. 
Mattern et al.\cite{MMJSSB23} further enhanced attack performance by introducing the neighborhood attack. 
In this approach, the adversary creates a ``neighborhood'' sample by replacing some words in the target sample. 
If the target sample exhibits significantly lower loss than its neighborhood sample, it is deemed more likely to be a member. 
Shi et al.~\cite{SAXHLBCZ23} proposed a conjecture suggesting that unobserved examples are more prone to contain outlier words with low probabilities under the Language Model (LLM), while a recognized example is less likely to include words with such diminished probabilities. 
They determine membership status based on the average likelihood of tokens with low probabilities.
However, all these attacks rely on access to the posterior or probability associated with the generated response.

While some research aims to infer membership status with label-only access, applying these attacks to the linguistic domain may prove challenging. 
On one hand, the input space of natural language lacks the continuity found in images. 
On the other hand, language models are typically much larger than vision models, introducing difficulties in crafting adversarial examples to measure the distance between the target sample and the decision boundary.

%-------------------------------------------------------------------------------
\section{Prompt Design}
\label{append:prompt}
%-------------------------------------------------------------------------------

The template of the prompt for different datasets can be found in~\autoref{tab:format1}.

\begin{table*}
\centering
\caption{Examples of the prompts used for text classification for the ICL setting.}
\begin{tabular}{p{1.0cm}p{7.6cm}p{4.2cm}}
\toprule
\textbf{Task} & \textbf{Prompt} & \textbf{Label Names} \\
\midrule
DBPedia & Classify the documents based on whether they are about a Company, School, Artist, Athlete, Politician, Transportation, Building, Nature, Village, Animal, Plant, Album, Film, or Book.
\vspace{3.9pt}\hspace{-0.1cm}

Article: Leopold Bros. is a family-owned and operated distillery located in Denver Colorado. \newline
Answer: Company
\vspace{3.9pt}\hspace{-0.1cm}

Article: Aerostar S.A. is an aeronautical manufacturing company based in Bacău Romania. \newline
Answer: &  Company, School, Artist, Athlete, Politician, Transportation, Building, Nature, Village, Animal, Plant, Album, Film, Book \\

\midrule
AGNews & Article: Kerry-Kerrey Confusion Trips Up Campaign (AP),"AP - John Kerry, Bob Kerrey. It's easy to get confused." \newline Answer: World
\vspace{3.9pt}\hspace{-0.1cm}

Article: IBM Chips May Someday Heal Themselves,New technology applies electrical fuses to help identify and repair faults. \newline Answer: &  World, Sports, Business, Technology \\

\midrule
TREC & Classify the questions based on whether their answer type is a Number, Location, Person, Description, Entity, or Abbreviation. \vspace{3.9pt}\hspace{-0.1cm}

Question: What is a biosphere? \newline
Answer Type: Description
\vspace{3.9pt}\hspace{-0.1cm}

Question: When was Ozzy Osbourne born? \newline
Answer Type: &  Number, Location, Person, Description, Entity, Abbreviation \\

\bottomrule
\end{tabular}
\vspace{-0.2cm}
\label{tab:format1}
\end{table*}

\end{document}